\documentclass[journal]{IEEEtran}
\usepackage{times}
\usepackage{epsfig}
\usepackage{color}
\usepackage{mysetup_subm}

\IEEEoverridecommandlockouts
\renewcommand{\QED}{\QEDopen} 
\newenvironment{myproof}{\begin{proof}}{\end{proof}\vspace{2mm}}

\newcommand{\refeqs}[2]{(\ref{#1}-\ref{#2})} 
\newcommand{\ve}{\varepsilon}
\renewcommand{\eps}{\ensuremath{\ve}}

\newcommand{\TP}{\ensuremath{T_{\text{P}}}} 

\newcommand{\MyDelta}{\text{{\scriptsize\mbox{$\mathbf{\Delta}$}}}}
\newcommand{\Deltat}{\ensuremath{\MyDelta t}}
\newcommand{\DeltaT}{\ensuremath{\MyDelta T}}

\newcommand{\epsS}{\ensuremath{\ve_{\mathrm{s}}}}
\newcommand{\deltaS}{\ensuremath{\delta_{\mathrm{s}}}}

\newcommand{\TwoFIFO}{two-priority FIFO}

\begin{document}

\title{Oscillations with TCP-like Flow Control in Networks of Queues\thanks{Preliminary version of this paper has appeared in \emph{IEEE Infocom 2006}. The current version (dated November 2005, with a minor revision in December 2008) is available from arXiv. }}

\author{Matthew Andrews~and~Aleksandrs~Slivkins
\thanks{M. Andrews is with Bell Laboratories, Lucent Technologies,
Murray Hill, NJ 07974, USA.
	Email: \emph{andrews at research.bell-labs.com}.}%
\thanks{A. Slivkins is with Microsoft Research, Mountain View, CA 94043.
	Email: \emph{slivkins at microsoft.com}.
This work has been completed while A. Slivkins was a graduate student at Cornell University. Parts of this work were done while A. Slivkins was a summer intern at Bell Labsoratories, Lucent Technologies.}%
}

\date{November, 2005}

\maketitle
\thispagestyle{plain}

\begin{abstract}
We consider a set of flows passing through a set of
servers.  The injection rate into each flow is governed by a flow
control that increases the injection rate when all the servers on the
flow's path are empty and decreases the injection rate when some
server is congested.  We show that if each server's congestion is
governed by the arriving traffic at the server then the system can
{\em oscillate}.  This is in contrast to previous work on flow control where
congestion was modeled as a function of the flow injection rates and
the system was shown to converge to a steady state that maximizes an
overall network utility. 
\end{abstract}

{\bf Categories and subject descriptors:}
F.2.2 [Nonnumerical Algorithms and Problems]: Sequencing and scheduling.

{\bf General Terms:} algorithms.

{\bf Keywords:} routing, scheduling, stability.

\section{Introduction}

In this paper we study how TCP-like mechanisms for flow control
operate in a network of queues with feedback.  We consider a set of
flows\footnote{Throughout the paper we use the terms {\em flow}
and {\em session} interchangeably.} passing through a network of
finite capacity servers.  Let $x_i(t)$ be the injection rate of data
into flow $i$. A series of papers,
(e.g.~\cite{Kelly97,KellyMT98,LowPW02,HollotMTG01,KunniyurS00,MassoulieR99,%
ChoeL03,LiuBS05,LowL99,MoW00}) has shown that many variants of TCP can
be viewed as solving optimization problems on the variables $x_i$
subject to the constraint that none of the server capacities are
overloaded.

To give a concrete example, consider the following model of
TCP-like additive-increase-multiplicative-decrease (AIMD) flow control
considered by Kelly et al.\ in \cite{KellyMT98}.  Let $\mu_{Q}$ be the
service rate of server $Q$.
Let $S(Q)$ be the set of flows that pass through server $Q$.
In \cite{KellyMT98}, the injection rate into flow $i$
is controlled by
\begin{eqnarray}
	\dot{x}_i(t) &=& \textstyle \kappa\left(
		w_i-x_i(t)\sum_{Q:i\in S(Q)}\gamma_{Q}(t) \right)
	\label{eq:Kelly1}\\
	\gamma_{Q}(t)&=& \textstyle p_{Q}\left(
		\sum_{i\in S(Q)}x_i(t)
	\right).\label{eq:Kelly2}
\end{eqnarray}
In the above equations $\kappa$ and $w_i$ are parameters,
$\dot{x}_i(t)$ represents $\frac{d}{dt}x_i(t)$, $\sum_{i\in
S(Q)}x_i(t)$ represents the current congestion at server $Q$ and
$p_{Q}(y)$ is a price charged by server $Q$ when the congestion is
$y$. An example given in \cite{KellyMT98} is
	$$ p_{Q}(y)=(y-\mu_{Q}+\ve)^{+}/\ve^2 $$
for some suitable parameter
$\ve$.  Kelly et al.\ show in \cite{KellyMT98} that all trajectories of the
above system converge to a stable point that maximizes the utility
function,
$$
U(x)=\sum_i w_i \log x_i -
\sum_{Q} \int_0^{\sum_{i\in S(Q)}x_i}p_Q(y)\,dy.
$$
For sufficiently small values of $\ve$ this is approximately equal
to solving the system,
$$
\textstyle \max \sum_r w_i\log x_i
$$
subject to:
$$
\textstyle \sum_{i\in S(Q)}x_i \le \mu_Q~~~~~\forall \ell.
$$
Since \cite{KellyMT98} appeared there has been much other work showing
that many other variants of TCP solve an appropriately defined utility
maximization problem.  However, a common feature of this work is that
the behavior of servers in the network is directly affected by the
{\em injection rates} of the flows that pass through the server. In
other words, the congestion at server $Q$ is modeled as a function of
$\sum_{i\in S(Q)} x_i(t)$. We refer to such models as {\em
injection-rate based} models.  These models are often motivated by
networks where congestion is noticed and signaled to the flow
endpoints before the link is saturated. In this case queues never
build up.

In most deployed networks however, congestion is only noticed by TCP
when a server is saturated and the associated queue does build up. In
this case the queue dynamics mean that the flow arrival rates at
internal servers in the network may be different than the external
flow rate.  In order to capture such situations, in this paper the
congestion will be a function of $\sum_{i\in S(Q)} a_{(i,Q)}(t)$, where
$a_{(i,Q)}$ is the {\em arrival rate} of flow $i$ at server $Q$ at time
$t$. We refer to a model that models congestion in this way as an {\em
arrival-rate based} model.  As already mentioned, the queueing
behavior in a network can mean that $a_{(i,Q)}(t)$ behaves dramatically
differently than $x_i(t)$. Indeed, a large body of work
(e.g.\cite{RybkoS92, DaiHV99, AndrewsAFKLL96, Andrews04-fifo, BhattacharjeeGL04})
has looked at instability phenomena in networks. In these phenomena,
the queueing behavior at the servers causes an unbounded amount of
data to build up in the network, while at the same time the injection
rates into the network do not saturate any link, i.e.\ $\sum_{i\in
S(Q)}x_i(t)< \mu_{Q}$ for all $t$ and $Q$. This means that even though
server capacities are never violated, queue occupancies and end-to-end
delays grow without bound.  The reason that these instabilities occur
is that the arriving traffic creates oscillations in the network.
Each queue alternates between periods when it is empty and periods
during which a lot of data arrives, during which the queue grows
large. These oscillations grow longer with time and so the queue sizes
grow larger with time.

In this paper we examine whether such oscillations can occur when the
injection rates are governed by a flow control such as
(\ref{eq:Kelly1})-(\ref{eq:Kelly2}), in the case when the congestion at a
link is governed by $\sum_{i\in S(Q)} a_{(i,Q)}(t)$ rather than
$\sum_{i:i\in S(Q)} x_i(t)$. In particular we consider a network of
servers, most of which serve data according to First-in-First-Out
(FIFO).  The rate at which data arrives at internal nodes of the
network is governed by the dynamics of the queues.  When data leaves
the network the amount of delay that the data has experienced within
the network affects the new injection rates into the network, in a
manner similar to (\ref{eq:Kelly1})-(\ref{eq:Kelly2}). Our main
result is that there exists an initial state of the network from which
the system has periodic oscillations. This in turn means that the flow
control never converges to an optimal set of injection rates. Moreover, the oscillations in our example are very wide: all injection, arrival and departure rates, and all queue heights, oscillate between 0 and their respective maximal values.

We stress that the oscillations in our example are solely due to the
queueing dynamics.  This provides a contrast to the oscillating
examples of Choe and Low~\cite{ChoeL03} and Liu et al.~\cite{LiuBS05} in which
oscillations occurred due to the introduction of propagation delays into
an injection-rate based model.  We discuss this distinction in more
detail in Section~\ref{s:remarks}.

\subsection{The model: network dynamics}

We consider a set of servers and a set of flows.  Each flow consists
of a path through the network. For ease of analysis we model data as a
continuous fluid.  We begin by describing how we model the network dynamics. We then describe how we model the flow control.

Let us consider a given server $Q$. We will use the following notation. We let $\mu$ be the service rate of $Q$, and $S$ be the set of flows that pass through $Q$. For a time $t$, we define $h(t)$ to be the height of server $Q$, which is the amount of fluid queued at server $Q$ normalized by the service rate. We also let $a_i(t)$ and $d_i(t)$ be the arrival and departure rates, respectively, of fluid on flow $i$ at server $Q$.

Most of the servers in our oscillating examples will be FIFO servers.
A FIFO server $Q$ is governed by the following equations (see e.g.\
\cite{Bramson96}).
\begin{eqnarray}
d_i(t+h(t)) &=& \frac{a_i(t)}{\dot{h}(t)+1}.\label{eq:fifo1}
\end{eqnarray}
If $h(t)>0$,
\begin{eqnarray}
	\dot{h}(t) &=& \textstyle \left(
		\frac{1}{\mu}\sum_{i\in S}a_i(t)
	\right)-1.\label{eq:fifo2}
\end{eqnarray}
If $h(t)=0$,
\begin{eqnarray}
	\dot{h}(t)&=&\textstyle \max\left\{\left(
		\frac{1}{\mu}\sum_{i\in S}a_i(t)
	\right)-1,\; 0\right\}.\label{eq:fifo3}
\end{eqnarray}
That is, any fluid that arrives at server $Q$ at time $t$ will depart
at time $t+h(t)$. The departure rate of this fluid will equal the
arrival rate divided by a quantity that is proportional to the
aggregate arrival rate.

When fluid leaves server $Q$ it goes to the next server on its path, call it $Q'$.  The arrival rate at $Q'$ is determined by the
departure rate from server $Q$ as well as the propagation delay
between the two servers, call it $\tau$. In particular we have
	$$a_{(i,Q')}(t)= d_{(i,Q)}(t-\tau). $$
In our construction all propagation delays will be zero.

\OMIT{
When fluid leaves server $Q$ it goes to the next server on its path,
$n_{(i,Q)}$.  The arrival rate at $n_{(i,Q)}$ is determined by the
departure rate from server $Q$ as well as the propagation delay
between the two servers. In particular we have
	$$a_{(i,n)}(t)= d_{(i,Q)}\left(t-\tau_{(Q,n)}\right),\;
		\text{where $n = n_{(i,Q)}$}$$
} 


In addition to the FIFO servers, a small number of servers in the
network divide the flows into two classes and give priority to the
first class over the second class; we call them \emph{\TwoFIFO} servers. For these servers the flows from
the first class behave as if they are being served by a FIFO queue but
the only arrivals come from the first class flows. The second class flows
are served according to FIFO among themselves but the service rate
available to them is limited to the amount of capacity that is not
used by the first class flows.

In Section~\ref{sec:allFIFO} we conjecture that these priorities are not necessary and that our oscillating example can be modified so all the queues are FIFO. The difficulty in proving this conjecture lies in arrival rates that
are negligibly small but cumbersome to analyze.

\renewcommand{\epsd}{\ensuremath{{(\ve,\delta)}}}
\newcommand{\refAS}[1]{(A\ref{#1})} 
\newcommand{\refASS}[2]{(A\ref{#1}-A\ref{#2})} 

\subsection{The model: parameterized flow control}
\label{sec:flowControl}

We now describe our flow control mechanism.  Consider some session $S$ at a
given time $t$. Say it is \emph{blocked} if at least one server $Q$ on its
path stores either a positive amount of session $S$ fluid, or a positive amount of some other session's fluid whose priority at $Q$ is higher than that of $S$. Say session $S$ is
\emph{happy} if any server $Q$ on its path does not store any session
$S$ fluid and, moreover, the bandwidth available to $S$ at $Q$ given
this server's priorities is strictly larger than the injection rate of
$S$ into $Q$. Say session $S$ is \emph{stable} if it is neither blocked nor happy, i.e. if it is bottlenecked at the current injection rate. We assume that the injection rate of $S$ should increase when $S$ is happy, decrease when $S$ is blocked, and stay constant when $S$ is stable.

For the purposes of this paper we will assume that for a given session the flow control mechanism is given by two functions,
	$f(\ve,t)$ and $g(t)$,
which control flow decrease and flow increase, respectively. Specifically:
\begin{itemize}

\item $f(\ve,t)$ is the injection rate at time $t$ assuming that at time $0$ the session becomes blocked and stays blocked till time $t$, and at time $0$ the injection rate is $\ve$.

\item $g(t)$ is the injection rate at time $t$ assuming that at time $0$ the injection rate is 0, and between times $0$ and $t$ the session is happy.

\end{itemize}

We will allow an arbitrary flow increase function $g(t)$ as long as it is monotone and increases to infinity. We consider two main examples of flow decrease: \emph{additive decrease}:
\begin{equation}\label{eq:additiveDecrease}
	f_\delta(\ve,t) = \ve - \delta t,
\end{equation}

\noindent and \emph{multiplicative decrease}
\begin{equation}\label{eq:multDecrease}
f_\delta(\ve,t) = \begin{cases}
	\ve e^{-\delta t} 		& t\leq t_0 \\
	\alpha\ve - \delta (t-t_0) 	& \text{otherwise}
\end{cases}
\end{equation}
\noindent where $\alpha\in(0,1)$ is a small constant and $t_0$ is the time such that
	$e^{-\delta t_0} = \alpha$.
We use~\refeq{eq:multDecrease} instead of the ``simple''  multiplicative decrease
	$f_\delta(\ve,t) = \ve e^{-\delta t}$
because for technical reasons we need the injection rate to drop to $0$ in some finite time.

\OMIT{ 
\begin{equation}\label{eq:multDecrease}
f_\epsd(t) = \begin{cases}
	\ve e^{-\delta t} 	& \text{$\eps>\Delta$ and $t\geq t_0$} \\
	\Delta - \delta (t-t_0) & \text{$\eps>\Delta$ and $t<t_0$} \\
	\eps-\delta t		& \text{otherwise}
\end{cases}
\end{equation}
} 

In both examples $\delta>0$ is the \emph{damping parameter} which controls the speed at which the flow rate decreases. We will assume that we are given a pair of functions $(f,g)$ which all servers must use, but we are allowed to tune $\delta$. To make the model slightly more general, we will assume that the flow increase function $g(t)$ is also parameterized by this $\delta$.

With the above types of flow control we can model both
additive-increase-additive-decrease (AIAD) and
additive-increase-multiplicative-decrease (AIMD) types of flow control.  We
recall that AIMD schemes have been widely studied in connection
with providing effective flow control in the internet.  (See e.g.\
\cite{ChiuJ89,Jacobson88}.)

\newcommand{\VALUESofDelta}{\ensuremath{S_{\text{{\sc del}}}}}

To generalize examples~\refeq{eq:additiveDecrease} and~\refeq{eq:multDecrease}, and to treat them in a unified model, we define the flow control mechanism as a triple
	$(f_\delta,g_\delta,\VALUESofDelta)$,
where $\VALUESofDelta\subset \R$ is the set of possible values for the damping parameter $\delta$, and the functions $(f_\delta, g_\delta)$ satisfy the following set of axioms:

\begin{enumerate}
\renewcommand{\labelenumi}{(A\arabic{enumi})}
\item \label{ass:increase}
\emph{increase to $\infty$}: $g_\delta(t)$ increases to infinity for any $\delta$.

\item \label{ass:decrease}
\emph{decrease to $0$}: for any fixed $\delta$ and $\ve$, function $f_\delta(\ve,t)$ becomes $0$ at some positive time, and then stays $0$.

\OMIT{ 
\item \label{ass:consistency}
\emph{consistency}: for any rates $0<\eps_1<\eps_2$ and a fixed $\delta$,
$f_{(\eps_1,\delta)}$ coincides with the appropriately delayed $f_{(\eps_2,\delta)}$:
	$$f_{(\eps_1,\delta)} (t) = f_{(\eps_2,\delta)} (t+ \Deltat)
		\text{ for all $t$,}$$
where $\Deltat$ is the (smallest) time such that $f_{(\eps_2,\delta)} = \eps_1$.
} 

\item \label{ass:monotonicity}
 \emph{monotonicity}: function
	$ f_\delta(\ve,t)$
is increasing in variable $\ve$ and decreasing in parameter $\delta$ and in variable $t$.

\item \label{ass:power}
\emph{damping}: for any fixed rates $0<\ve'<\ve$ and any fixed time $t$, there exists some $\delta$ such that
	$f_\delta(\ve,t) > \ve'$,
and there exists some other $\delta$ such that
	$f_\delta(\ve,t) < \ve'$.

\item \label{ass:continuity}
\emph{continuity}: for any fixed $\delta$ the function
	$f_\delta(\ve,t)$
 is a continuous function from $\R^2$ to $\R$.

\OMIT{  
\item \label{ass:halving}
\emph{halving the rate}: for any fixed rates $0<\ve_1< \ve_2$ and any fixed time $t$ there exists $\delta$ such that for any initial rate $\ve\in [\ve_1; \ve_2]$ we have
	$f_\delta(\ve,t)\geq \ve/2$.
} 
\end{enumerate}

Axioms~\refASS{ass:increase}{ass:monotonicity} are very natural. Axiom~\refAS{ass:power} expresses why it is useful to have a damping parameter. Finally, Axiom~\refAS{ass:continuity} captures the intuition that in the fluid-based model everything ought to be continuous. All five axioms easily satisfied by additive  decrease~\refeq{eq:additiveDecrease} and multiplicative decrease~\refeq{eq:multDecrease}.

Axiom~\refAS{ass:increase} is the only axiom on flow increase; note that it says nothing about the damping parameter. We will tune $\delta$ solely to obtain the desired flow decrease. In particular, for flow~\emph{increase} we will tolerate an arbitrary dependence on $\delta$.

Note that the function $f_\delta(\ve,t)$ does not need to be continuous in $\delta$, and the set \VALUESofDelta\ of possible values of $\delta$ does not even need to be a continuous real interval. For example, both for additive decrease~\refeq{eq:additiveDecrease} and for multiplicative decrease~\refeq{eq:multDecrease} this set can be discrete, e.g.
	$\VALUESofDelta =   \N \cup \{ 1/n:\, n\in\N \}$.

\subsection{The main result and extensions}
\label{sec:mainResult}

A \emph{network of servers and sessions} is a network of servers connected by server-to-server links, together with a set of sessions; here each session is specified by a source-sink pair, a flow path, and a flow control mechanism.
At a given time the \emph{state} of such network includes:
\begin{itemize}
\item injection rates into each session,
\item height and composition of each queue
\end{itemize}
Here the composition of a given queue $Q$ includes, for each height
$h$ and each session $S$ passing through $Q$, the density of session
$S$ fluid at height $h$. (This density is defined to be the rate at
which session $S$ fluid leaves queue $Q$ after time $h$.) We say that
a network \emph{oscillates} if starting from some state it eventually
reaches this state again.

We say that at a given time the injection rates are \emph{feasible} if they do not overload any server: for each server $Q$, the sum of the current injection rates of all sessions that go through $Q$ is no bigger than the service rate of $Q$.

\OMIT{A state of a network is called \emph{feasible} if the current injection rates are feasible.}

Our main result is that given any parameterized flow control mechanism, we can create a network that exhibits wide oscillations during which the injection rates stay feasible. We state this result as follows:

\begin{theorem} \label{thm:main}
Suppose we are given an arbitrary flow control mechanism that satisfies
	properties~\refASS{ass:increase}{ass:continuity},
and we are allowed to choose the damping parameter separately for each session. Suppose all servers must be either FIFO or \TwoFIFO, and we are allowed to choose the service rate separately for each server.

Then there exists a network of servers and sessions that {\sc oscillates} so that the injection rates are feasible at all times. The damping parameters take only three distinct values.

The oscillations are very wide, in the sense that all injection, arrival and departure rates, and all queue heights, oscillate between 0 and their respective maximal values.
\end{theorem}

It is interesting to ask if we can fine-tune the above result so that all sessions are really running the same flow control mechanism, i.e. have the same damping parameter. We can do it for the case of additive decrease~\refeq{eq:additiveDecrease} and more generally for all flow decrease functions of the form
\begin{equation}\label{eq:splittableExample}
\text{
	$f_\delta (\ve, t) = \ve - \delta\cdot\ve^\alpha\cdot t^\beta$,
	where $\alpha<1$ and $\beta>0$.
}\end{equation}
Note that any such function satisfies axioms~\refASS{ass:increase}{ass:continuity}. In particular, by differentiating we can see that it is increasing in $\ve$ whenever it is non-negative.

\begin{theorem} \label{thm:sameDelta}
Suppose in Theorem~\ref{thm:main} the flow decrease function satisfies~\refeq{eq:splittableExample}. Then we can choose the same damping parameter for all sessions. Moreover, there exists $\delta^*\in\VALUESofDelta $ with the following property: for any
	$\delta\in\VALUESofDelta$ such that $\delta\leq \delta^*$
we can choose the damping parameter to be $\delta$.
\end{theorem}

In fact, our result applies to a somewhat more general family of flow decrease functions, see Section~\ref{sec:sameDelta} for further discussion. Note that we can take any given $\delta$ as the common damping parameter, as long as it is sufficiently small.

\OMIT{ 
\begin{theorem} \label{thm:AD}
Suppose the flow decrease function $f_\delta$ satisfies
\begin{equation}\label{eq:parallelFlows}
	f_\delta(\ve, t) = n\cdot f_{\delta/n}(\ve/n, t)
\end{equation}
for any parameter $\delta\in \VALUESofDelta$, any time $t\geq 0$ and any positive integer $n$. Moreover, assume that in axioms~\refAS{ass:power} and~\refAS{ass:halving} we can choose
	$\delta\in\mathbb{Q}$.
Then in Theorem~\ref{thm:main} we can choose the same damping parameter for all sessions.
\end{theorem}
} 

\subsection{Remarks}
\label{s:remarks}

\OMIT{ 
We remark that our oscillating example can be extended to the case of
TCP Vegas-type schemes~\cite{BrakmoP95,LowPW02,ChoeL03} where the flow rate is
increased/decreased at a rate of $1/(\Gamma_i(t))^2$ depending on
whether or not there is congestion on the path of flow $i$.  Here
$\Gamma_i(t)$ is the end-to-end-delay on the path of flow $i$.  The
only difference from the proof of Theorem~\ref{thm:main} is that
we need to introduce propagation delays in the example in order to
make sure that $1/(\Gamma_i(t))^2$ is always bounded. We omit the
details.
}

We stress that in our example it is the {\em dynamics of the queues}
that cause the oscillations.  This provides a contrast with the work
of Choe and Low~\cite{ChoeL03} and Liu et al.~\cite{LiuBS05} that
showed that injection-rate based models of TCP Vegas and a variant
called Stabilized Vegas can both be unstable when feedback delays are
present.  These papers model TCP as a set of differential equations
with feedback and show that the feedback delays can cause the
equations to oscillate.  In particular, for each server $Q$ the
injection rates of flows in $S(Q)$ oscillate between values that
overload $Q$ and values that underload $Q$.  In contrast, in our
oscillating examples the injection rates $x_i(t)$ are such that
$\sum_{i\in S(Q)} x_i(t)\le \mu_Q$ always holds and so the injection
rates do not {\em a priori} overload any server. The oscillations
arise because the queueing dynamics mean that temporary congestion is
continually being created in the network.

Another feature of our model is that when congestion occurs the
injection rate into a session decreases {\em continuously}. Some
previous work has looked at a contrasting model of TCP in which the
window size (and hence the injection rate) of a session is
instantaneously decreased by half whenever congestion occurs.
Baccelli et al.\cite{BaccelliMR02} showed that these ``jumps'' can lead
to oscillations when a system of TCP sessions interacts with a RED or
drop-tail queue. Our example shows that even if these jumps are
eliminated using a continuous decrease then the flow control can still
oscillate when interacting with a network of servers.

One paper that considers a similar problem to ours is Ajmone Marsan
et al.~\cite{Marsan-ton04}. They present an example consisting of a network of
finite-buffer servers in which the scheduling discipline is either
strict-priority or Generalized Processor Sharing. The example is
constructed in such a way that if the source rates are non-adaptive
then the queue sizes will build up in an oscillating manner. The
paper~\cite{Marsan-ton04} demonstrates via simulation that if we instead use
adaptive additive-increase-multiplicative-decrease sources then we
still get queue buildups that lead to packet losses and oscillating
behavior. There are a number of differences between the model of~\cite{Marsan-ton04} and our model however. The main difference is that~\cite{Marsan-ton04} uses
finite buffers and the sources only adapt to packet losses. In our
model we consider unbounded buffers but the sources respond directly
to congestion. Our results therefore demonstrate that as long as
sources respond to congestion on a link, we can still get oscillating
behavior and hence suboptimal utilization, even if no packet losses
occur. Another difference between the two models is that~\cite{Marsan-ton04} uses
networks of strict-priority or GPS servers whereas our servers are
mostly FIFO.

Finally we remark that one of the main reasons that we are able to
create oscillations is that the queueing disciplines at the servers
are oblivious to the state of the flow control.  In contrast, there
exist schemes in which the flow control protocols and the queueing protocols
work in combination and are able to ensure convergence and prevent
oscillations. See for example the \emph{Greedy primal-dual algorithm} of
Stolyar~\cite{Stolyar05-gpd}.

\subsection{Preliminaries}

\OMIT{ 
For the case of additive decrease~\refeq{eq:additiveDecrease} we take
an advantage of the fact that a single session with parameters
$(\ve,\delta)$ decreases its rate in exactly the same way as $n$
parallel sessions with parameters $(\ve/n,\delta/n)$. We modify the
main construction so that for all sessions the damping parameters are rational, and then split the sessions so that all of the resulting sessions have the same damping parameter.
We defer the details to the \full.
} 

Without loss of generality we may assume that we can choose the maximal injection rate for a given session. Indeed, for any target maximal rate $\ve$ we can attach a new server in the very beginning of the session path, with a service rate $\ve$. For simplicity, in the following sections each session will have \emph{two} flow control parameters $(\ve,\delta)$, where $\ve$ is the maximal injection rate and $\delta$ is the damping parameter.

In practical settings it makes no sense to have sessions with
non-simple flow paths (i.e. flow paths that go through the same server
more than once). However, non-simple flow paths are often useful
theoretically since they can make a construction more compact and/or
clear. It turns out that one can use non-simple flow paths without
loss of generality, since any network with non-simple flow paths can
be converted to an equivalent network where all flow paths are simple,
e.g. see \cite{Andrews05-fifo}. The basic idea is that if $Z$ is the
maximum number of times that a session visits a server then we create
$Z$ copies of the network. Whenever a session is about to visit a
server for second time it simply moves to a new copy of the
network. More precisely, we organize the copies into a ring: we number the copies from $0$ to $Z-1$, and the next visit from copy $i$ goes into copy $(i+1) \pmod{Z}$. In this way we can create an example in which each session
visits a server at most once.  However, in order to avoid this extra
complexity, we will use non-simple flow paths without any further
notice.

\subsection{Organization of the paper}

In Section~\ref{sec:review} we overview our construction. In Section~\ref{sec:gadget} we describe the main building block in our construction, which we call \emph{the basic gadget}. Then in Section~\ref{sec:fullConstruction} we proceed to the full construction and the proof of its performance. In Section~\ref{sec:sameDelta} we discuss the extension where we choose the same damping parameters for all session. In Section~\ref{sec:allFIFO} we conjecture an extension where all servers are FIFO. Finally, in Section~\ref{sec:conclusions} we conclude and state some open questions,

\section{Overview of Oscillating Example}
\label{sec:review}

The complete description of our oscillating example is somewhat
complex and so we begin with a brief overview. Our construction is
based on rows of {\em gadgets}. Each row of gadgets has a single
horizontal session $H$ that passes through all the servers in the row
multiple times. Each gadget in a row consists of four servers $Q_1$,
$Q_2$, $Q_3$ and $Q_4$. (See Figure~\ref{fig:basicGadget}).  The
server $Q_4$ in one gadget is identified with the server $Q_1$ in the
next gadget.  The main aim of the gadget is to transfer session $H$
fluid from server $Q_1$ to server $Q_4$.  In order that this can occur
session $H$ loops multiple times through all of servers $Q_1$--$Q_4$
and then loops multiple times through server $Q_4$ only. We also have
a session $S$ that goes through server $Q_4$ only.

Initially, server $Q_1$ contains session $H$ fluid and server $Q_4$ is
empty. This means that session $S$ is not blocked and so it is injecting at
its maximum rate.  As session $H$ fluid loops through server $Q_4$ we
get a buildup in server $Q_4$.  This causes the injection rate of
session $S$ to decrease.  Eventually we reach a state in which all of
the session $H$ fluid is now in $Q_4$, the injection rate of session
$S$ is zero, and there is no session $S$ fluid in server $Q_4$.

Since server $Q_4$ in the current gadget is identified with server $Q_1$
in the next gadget we can repeat this process.  Note that during the
process the injection rate of session $S$ has decreased from its
maximum rate to zero. Once the session $H$ fluid has left server $Q_4$
and moved to the next gadget, the injection rate of session $S$
increases again to its maximum rate.  Hence we have created {\em
oscillations} in the injection rate of session $S$.

Unfortunately, we cannot continue the above process indefinitely.
This is because in a finite network the session $H$ has finite length
and so the session $H$ fluid will eventually reach its destination.
We therefore need a mechanism to {\em replenish} the session $H$
fluid. We do this as follows.

At the beginning of the row of gadgets we have a server called a
replenishing server. Initially all the servers in the row gadgets are
empty and so session $H$ fluid is injected at its maximum rate.  We
are able to create a buildup of fluid in the replenishing server.
This causes session $H$ fluid to build up in the replenishing server
and it also causes the session $H$ injection rate to go to zero.
When enough session $H$ fluid has been collected it begins the process
of passing through the row of gadgets.

It remains to describe how we create a buildup of fluid in the
replenishing queue. We do this in the following manner.  As the
session $H$ fluid passes through the server $Q_2$ in a gadget, it
creates a buildup at server $Q_2$.  We have a new vertical session $V$ that
also passes through server $Q_2$.  Session $V$ is blocked by session
$H$ at server $Q_2$ and so we get a buildup of session $V$ fluid.  Our
complete example consists of multiple rows of gadgets.  Each vertical
session $V$ passes through multiple rows and multiple copies of
$Q_2$. Moreover, the session $V$ fluid gets blocked at each copy of
$Q_2$.  Each session $V$ eventually passes through a replenishing
queue at the beginning of a row of gadgets.  Since the session $V$
fluid has been blocked at multiple copies of $Q_2$ it enters the
replenishing queue at a high rate.  This causes the required buildup
in the replenishing queue.

The vertical sessions $V$ are depicted in Figure~\ref{fig:grid}. The
interaction of the vertical sessions with the replenishing server is
depicted in Figure~\ref{fig:superRows}.  We remark that each server
processes data in FIFO order with two exceptions.  First, each server
$Q_2$ gives strict priority to session $H$ fluid over session $V$
fluid.  Second, each replenishing server gives priority to session $V$
fluid over session $H$ fluid.  As we discuss in Section~\ref{sec:allFIFO}, we made a step towards relaxing the first requirement: we are able to create a (more complex) basic gadget where the servers $Q_2$ are strictly FIFO.  We are unfortunately unable to construct an example where all servers are strictly FIFO. In Section~\ref{sec:allFIFO} we conjecture that this is possible and we briefly discuss why we believe this is so.

\section{Basic gadget}
\label{sec:gadget}

In this section we'll define a basic gadget that will be used as a main building block in our construction. This gadget consists of four servers called $Q_1$, $Q_2$, $Q_3$ and $Q_4$, and two sessions. We'll describe a phase in which we transfer fluid from server $Q_1$ to server $Q_4$.  During this phase the height of server $Q_2$ will increase and will then decrease again. The behavior of server $Q_2$ will allow us (in the full construction) to eventually accumulate new fluid that we use to replenish fluid that eventually reaches its destination.

\subsection{Basic gadget: construction}
\label{sec:gadgetConstruction}

The gadget consists of the four servers $Q_1 \ldots Q_4$, and two sessions: one \emph{horizontal} session that we call $H$, and one \emph{simple} session that we call $S$. The horizontal session $H$ starts and ends outside the gadget; in fact, in the full construction one such session comes through multiple gadgets. For the simple session $S$, the source and the sink lie inside the gadget.

\begin{figure}
\centering
\includegraphics[width=3.2in]{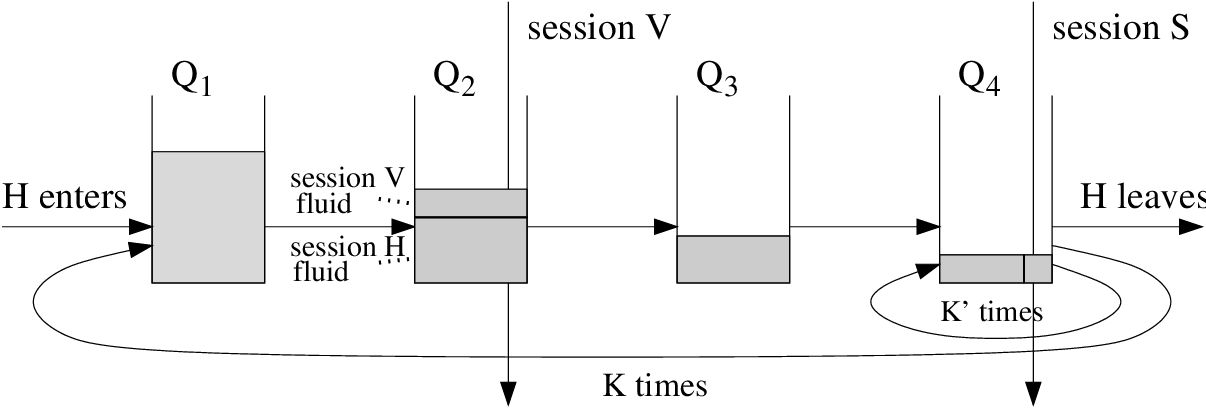}
\caption{A basic gadget. At server $Q_2$, session $H$ has priority
over session $V$. At server $Q_4$, sessions $S$ and $H$ are served
in FIFO order.}
\label{fig:basicGadget}
\end{figure}

Session $H$ enters the gadget at server $Q_1$ and leaves it at server $Q_4$. The path of $H$ first goes $K$ times through the \emph{big loop}, which consists of servers $Q_1$, $Q_2$, $Q_3$ and $Q_4$ in this order. The path then goes $K'$ more times through the \emph{small loop}, which consists of server $Q_4$ only. Here $K$ and $K'$ are parameters that we'll tune later. The path of the simple session $S$ goes through server $Q_4$ only. The two sessions compete in server $Q_4$ in FIFO order. See Fig.~\ref{fig:basicGadget} for the composition of a single basic gadget.

\begin{figure}
\centering
\includegraphics[width=3.2in]{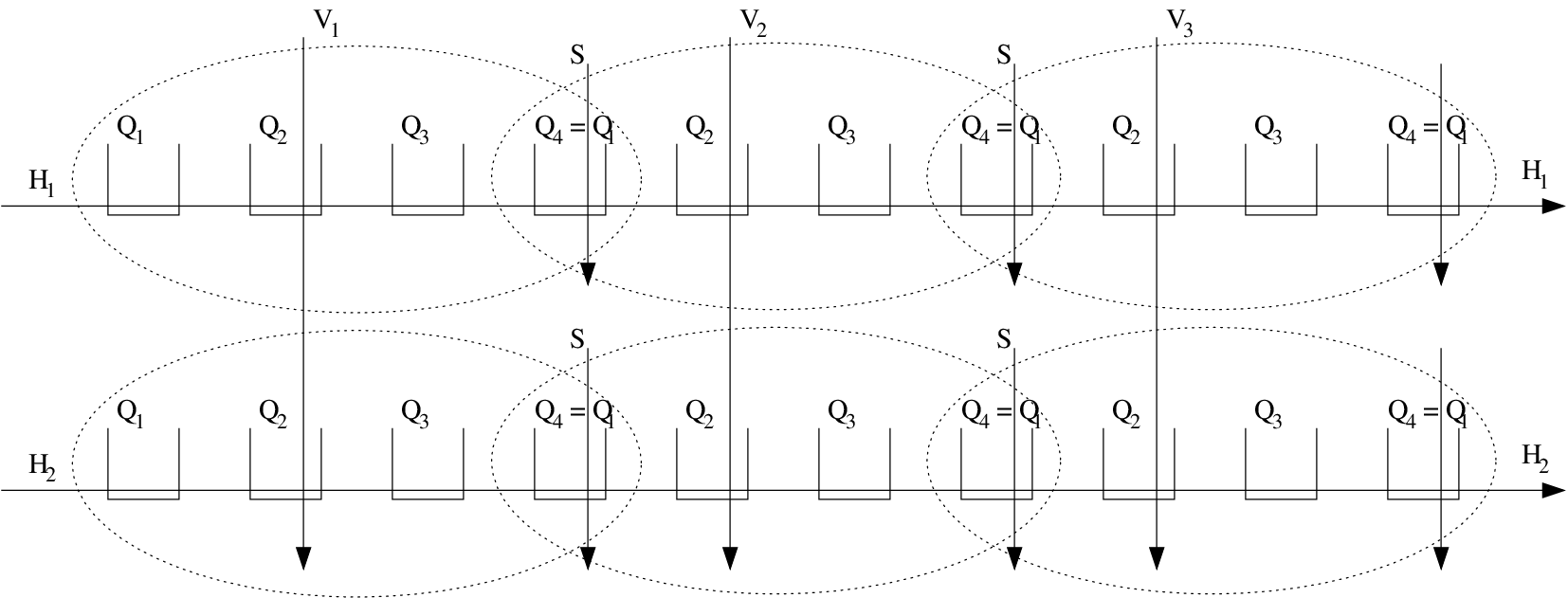}
\caption{The grid-like structure}
\label{fig:grid}
\end{figure}

Let us briefly preview how the basic gadgets fit into the full construction. There we shall have rows of servers that consist of many blocks of three servers each. Servers $Q_1$, $Q_2$ and $Q_3$ constitute one such block; server $Q_4$ will be the first server of the next block. The horizontal session $H$ goes consecutively through all blocks in a given row (more precisely, $H$ enters the next basic gadget right after it leaves the previous one; the details are Section~\ref{sec:fullConstruction}). The blocks of servers are also organized in columns; there will be \emph{vertical} sessions that go through consecutive servers $Q_2$ in the same column. These flows will not mix with the horizontal session $H$ since server $Q_2$ will strictly prefer $H$. This grid-like construction is summarized in Fig.~\ref{fig:grid}.

Now let us go back to the level of a single basic gadget.  We shall have several parameters (apart from $K$ and $K'$) that we'll tune later. For the simple session $S$, let $\epsS<1$ be the maximal flow rate, and let $\deltaS$ be the damping parameter. The flow control parameters of session $H$ are irrelevant at this point, simply because its source is outside of the scope. Moreover, throughout this section we'll assume that the injection rate of session $H$ into the gadget is 0. Servers $Q_1$ and $Q_4$ both have service rate $1$. Server $Q_2$ has service rate $1-\mu_2$ and server $Q_3$ has service rate $1-\mu_3$, for some parameters
	$0<\mu_2<\mu_3$.
We summarize the parameters in Table~\ref{tab:paramsGadget}.

\begin{table}
\renewcommand{\arraystretch}{1.3}
\centering
\begin{tabular}{l|l}
Parameter & Description   \\ \hline
$(\epsS,\deltaS)$ 	& flow control parameters for session $S$ \\
$K$		& number of big loops for session $H$\\
$K'$		& number of small loops for session $H$\\
$1-\mu_2$	& service rate of server $Q_2$ \\
$1-\mu_3$	& service rate of server $Q_3$
\end{tabular}

\caption{Tunable parameters in the basic gadget}
\label{tab:paramsGadget}
\end{table}

\subsection{Basic gadget: a single time phase}

In this subsection we describe the workings of a single time phase where we transfer one unit of session $H$ fluid from server $Q_1$ to server $Q_4$.

Let us define \emph{fresh} fluid as the fluid from session $H$ that has never come through server $Q_1$ within our basic gadget, i.e. as the session $H$ fluid that  still needs to go $K$ times through the big loop. Also, let us define \emph{processed} fluid as the fluid from session $H$ that either has come through our basic gadget, or just needs to come through server $Q_4$ exactly once to exit the gadget.

At the beginning of the phase we assume that server $Q_1$ contains one unit of fresh fluid, and the three other servers are empty. The injection rate of session $S$ is at its maximal value $\epsS$. Moreover, we assume that throughout the phase, the injection rate of session $H$ is zero. We finish with one unit of processed fluid at server $Q_4$.

The behavior of the servers during the phase is summarized in
Table~\ref{tab:phase}. During the time interval $[0,T_1]$, the height
of server $Q_1$ decreases and the height of servers $Q_2$, $Q_3$ and
$Q_4$ increases. During the time interval $[T_1,T_2]$, server $Q_1$ is
empty, the height of server $Q_2$ decreases and the height of servers
$Q_3$ and $Q_4$ increases.  During the time interval $[T_2,T_3]$,
servers $Q_1$ and $Q_2$ are empty,the height of server $Q_3$ decreases
and the height of server $Q_4$ increases. We ensure that
$T_3 = K/(1-\mu_3)$ and that $\ve(T_3) \geq \epsS/2$.

\newcommand{\UP}{\ensuremath{\uparrow}}
\newcommand{\DN}{\ensuremath{\downarrow}}

\begin{table}
\renewcommand{\arraystretch}{1.3}
$$\begin{array}{l|cccc|l}
\text{time $t$} & h_1	& h_2	& h_3	& h_4	& \text{remarks} \\ \hline
t=0   		& 1 	& 0 	& 0 	& 0 	&
				\text{fresh fluid only; $\ve(0) = \epsS$} \\
t\in [0,   T_1]	& \DN 	& \UP 	& \UP 	& \UP	&  \\
t\in [T_1, T_2]	& 0 	& \DN 	& \UP 	& \UP	&  \\
t\in [T_2, T_3]	& 0 	& 0 	& \DN 	& \UP	&
		T_3 = K/(1-\mu_3);\; \ve(T_3) \geq \epsS/2 \\
t\in [T_3,   \TP]	& 0 	& 0 	& 0 	& \geq 1&  \\
t=\TP		& 0 	& 0 	& 0 	& 1	&
				\text{processed fluid only; $\ve(\TP) = 0$}
\end{array}$$
{\sc note:} we assume that the injection rate of session $H$ is 0 for all $t$.\\
{\sc notation:} $h_i(t)$ is the height of server $Q_i$ at time $t$; $\ve(t)$ is the injection rate of session $S$ at time $t$; a vertical (horizontal) arrow means that a given quantity is strictly increasing (decreasing) with time. \\
{\sc note:} session $H$ arrives at server $Q_2$ at rate at most $1-\mu_3$ between times $T_1$ and $T_3$, and at rate $0$ between times $T_3$ and $\TP$.
\caption{Basic gadget: main events within a single phase}
\label{tab:phase}
\end{table}

We state our result as follows:

\begin{lemma}\label{lm:phase}
Consider the basic gadget with a given flow decrease mechanism for session $S$ that satisfies
	properties~\refASS{ass:decrease}{ass:continuity}.
 Assume that the injection rate into session $H$ is 0 at all times. Furthermore, suppose that at time $t=0$ server $Q_1$ contains one unit of fresh fluid, the three other servers are empty, and the injection rate into session $S$ is at its maximal value $\epsS$.

Then  there exist parameters $(\epsS, \deltaS, \mu_2, \mu_3, K')$ and times $(T_1, T_2, T_3,\TP)$ such that the gadget functions as shown in Table~\ref{tab:phase}. Moreover, this holds for any $K\geq 20$, in which case we can choose
	$\epsS \in (5/3K; 20/K)$
and set $\mu_3 = 2\epsS/5$.
\end{lemma}

Moreover, it turns out that after the parameters $(K,\epsS,\mu_3)$ are
chosen in the above fashion, one can pick \emph{any} $\mu_2\in (0,
\mu_3)$. This enables us to fine-tune such quantities as $T_2$ and the
maximal height of server $Q_2$. Although we do not use this feature in
the present proof, it is useful in the all-FIFO setting; see Section~\ref{sec:allFIFO} for details.

\subsection{Basic gadget: proof of the main lemma}
\label{sec:basicGadgetProof}

Let $h_i(t)$ be the height of server $Q_i$ at time $t$, and let $\ve(t)$ be the injection rate of session $S$ at time $t$.

We set $\mu_3 = 2\epsS/5$ and assume $0<\mu_2<\mu_3$. At first session $H$ fluid is served by server $Q_1$ at rate $1$. Since $\mu_2>0$, we start to get a buildup in server $Q_2$.  Session $H$ fluid leaves server $Q_2$ at rate $1-\mu_2$ which is larger than $1-\mu_3$, so we start to get a buildup in server $Q_3$, too. Then session $H$ fluid leaves server $Q_3$ at rate $1-\mu_3$ and arrives at server $Q_4$. At the beginning of the phase fluid also arrives to server $Q_4$ on session $S$ at rate close to $\epsS$. Since $1-\mu_3+\epsS>1$ we get a buildup at server $Q_4$.

As server $Q_4$ is congested, $\ve(t)$ starts to decrease; however, we'll make sure that it stays large enough -- say, at least $\eps/2$ -- until time $T_3$ when server $Q_3$ empties. Since
	$1-\mu_3+\epsS/2>1$,
it follows that (at least) until this time $h_4(t)$ keeps increasing. Moreover, until time $T_3$ session $H$ fluid loops back to server $Q_1$ after being served by $Q_4$ at rate less than $1-\mu_3$ (since $Q_4$ is also serving session $S$). In particular, $h_1(t)$ decreases until it becomes $0$.

To summarize, initially the height of server $Q_1$ decreases, and the heights of the three other servers increase. At some time $t=T_1$, server $Q_1$ empties.  At this point, the fluid that loops back from server $Q_4$ passes directly through server $Q_1$ and arrives at server $Q_2$ at a rate less than $1-\mu_3 $, which is less than $1-\mu_2$.  Therefore at time $T_1$ the height of server $Q_2$ starts to decrease, until at some time $t=T_2$ it empties. At this point session $H$ fluid starts to enter server $Q_3$ at a rate smaller than $1-\mu_3$, so its height starts to decrease.  We use $T_3$ to denote the first time that server $Q_3$ empties.  At this point {\em all of the session $H$ fluid is stored in server $Q_4$}.

By definition of time $T_3$, at any time $t\in (0, T_3)$ session $H$ fluid leaves server $Q_3$ at the rate $1-\mu_3$, so exactly $(1-\mu_3)t$ units of session $H$ fluid are served by $Q_3$ between time $0$ and $t$. Since only $K$ units of session $H$ fluid are available (more precisely, one unit that passes through the big loop $K$ times), it follows that $T_3 \leq T$, where $T = K/(1-\mu_3)$.

From now on, let us fix parameter $K$ and vary parameter $\epsS$. Define constants $\ve_1 = 5/3K$ and $\ve_2 = 20/K$. Later we will choose $\epsS$ from the interval $[\ve_1; \ve_2]$.

\begin{claim}\label{cl:chooseDelta2K}
There exists a positive constant $\delta^*$ such that
\begin{equation}\label{eq:chooseDelta2K}
	\ve(2K) = f_{\delta^*}(\ve, 2K)
		\geq \ve/2\; \text{for any $\ve \in [\ve_1; \ve_2]$}.
\end{equation}
\end{claim}

\begin{myproof}
Let us denote $\alpha_i = \ve_1\, (3/2)^i$ and let us cover the interval $[\ve_1; \ve_2]$ with smaller intervals
	$[\alpha_i; \alpha_{i+1}]$.
For each such interval let us choose $\delta = \delta_i$ such that
	$f_\delta(\alpha_i, 2K) \geq \alpha_{i+1}/2$;
such $\delta$ exists by Axiom~\refAS{ass:power} since
	$\alpha_{i+1}/2 < \alpha_i$.
Then~\refeq{eq:chooseDelta2K} is satisfied for any
	$\delta\geq \delta_i$
and $\ve\in [\alpha_i; \alpha_{i+1}]$. Finally, let us choose $\delta^*$ to be the smallest of the $\delta_i$'s.
\end{myproof}

Let us set the damping parameter $\deltaS$ to $\delta^*$ from Claim~\ref{cl:chooseDelta2K}. Note that since $\mu_3 = 2\epsS/5$ it follows that $T<2K$, so
\begin{equation}\label{eq:chooseDelta}
	\ve(T) \geq \epsS/2\; \text{for any $\epsS \in [\ve_1; \ve_2]$}.
\end{equation}

\OMIT{ 
 is an increasing function of parameter $\ve$.  Let $T'$ be the value of this function for $\ve = \ve_2$; note that unlike time $T$, time $T'$ is a constant. Now let us use~\refAS{ass:halving} to choose a positive $\delta$ such that $\ve(T') \geq \ve/2$ for all $\ve \in [\ve_1; \ve_2]$. Then~\refeq{eq:chooseDelta} follows since $T\leq T'$.

Recall that $\mu_3 = 2\ve/5$; for any given values of $\ve$ and $K$, we choose $\delta$ such that $\ve(T) = \eps/2$ (such $\delta$ exists by continuity). In other words, at the cost of letting $\delta$ be a function of $\ve$ and $K$, we ensure that $\ve(T) = \ve/2$ and, consequently, that $\ve(T_3)\geq \ve/2$ as required.
} 

We choose parameter $\epsS$ carefully to ensure that $T_3 = T$, or, equivalently, that at time $t=T_3$ all of the session $H$ fluid has passed through servers $Q_1$, $Q_2$ and $Q_3$ exactly $K$ times. More precisely, we have the following claim whose proof is deferred to Section~\ref{sec:claimEpsk}:

\begin{claim}\label{cl:epsk}
For any $K\geq 20$ there exists $\epsS\in (5/3K; 20/K)$ such that for any $\mu_2\in (0,\mu_3)$  we have $T_3 = K/(1-\mu_3)$.
\end{claim}

In the remainder of the phase, all of the session $H$ fluid has to do is loop through server $Q_4$ another $K'+1$ times, which takes time at least $K'+1$. For any given $K$ (and hence $\epsS$, $\deltaS$ and $T_3$), we choose parameter $K'$ to be the smallest integer such that by the time $T_3+K'$ the injection rate of session $S$ has dropped to zero, and all of the session $S$ fluid has left the system. Clearly, such $K'$ exists for any given $K$.

The phase ends at some time $t=\TP$ when all of the session $H$ fluid has exactly one loop to go in $Q_4$. Obviously, at this time all of this fluid is still queued in $Q_4$; since all of the session $S$ fluid has already left the system, $h_4(\TP)=1$ as required at the end of the phase.

To complete the proof of Lemma~\ref{lm:phase}, it remains to prove Claim~\ref{cl:epsk}.

\OMIT{
It only remains to make sure that the maximal height of server $Q_2$ is
	$h_2(T_1) = \TP/\cel{5\TP}$.
Note that we are still free to tune parameter $\mu_2$ in the range $(0,\mu_3)$.

\begin{claim}\label{cl:mu2}
Fix any $K\geq 12$ and the corresponding $\epsS$. Then there exists
	$\mu_2\in [\epsS/60; \epsS/5]$
such that $h_2(T_1) = \TP/\cel{5\TP}$. Moreover, time $T_1$ lies in the interval
	$[1/\epsS; 3/\epsS]$.
\end{claim}

This completes the proof of Lemma~\ref{lm:phase}; we defer the proofs of the two claims to the next section.

\section{Basic gadget: detailed computations}
\label{sec:gadgetDetails}

This section is devoted to detailed computations which lead to Claim~\ref{cl:epsk} and
Claim~\ref{cl:mu2} and thus complete the proof of Lemma~\ref{lm:phase}.

}

\subsection{Basic gadget: proof of Claim~\ref{cl:epsk}}
\label{sec:claimEpsk}

Let us keep $K$ fixed and assume that $\epsS \in [5/3K; 20/K]$. Recall that $\ve(t)$ is the injection rate into session $H$ at time $t$, and $h_i(t)$ is the height of server $Q_i$ at time $t$. Recall that we use $T_i$ to denote the first time that server $Q_i$ empties:
	$$T_i = \min \{ t>0:\, h_i(t) = 0 \}.$$
We will also use $d_i(t)$ to denote the rate at which session $H$ fluid departs from server $Q_i$ at time $t$.

A key quantity in our analysis is the total amount of session $H$ fluid queued in servers $Q_1$, $Q_2$ and $Q_3$. We use $h^{+}(t)$ to denote this quantity at time $t$.  We note that $h^{+}$ behaves like the amount of fluid in a single queue that is fed at rate $d_4(t)$ and has service rate $1-\mu_3$. Since $\ve(T_3)\geq \epsS/2$, by the analysis in the previous section server $Q_3$ drains \emph{after} servers $Q_1$ and $Q_2$. It follows that
	$$T_3 = \min \{ t>0:\, h^+(t) = 0 \} $$
and, in particular, $h^+(T_3) = 0$.

For any time $t\in [0; T_3]$ the following equations hold:
\begin{eqnarray}
d_3(t)	&=& 1-\mu_3	\label{eq:systemA}\\
h_4(t)	&=& \int_0^t (d_3(t)+\ve(t)-1)\; dt \label{eq:systemB} \\
h^{+}(t)&=& 1+\int_0^t (d_4(t) - 1 +\mu_3)\; dt  \label{eq:systemC} \\
d_4(t+h_4(t))
	&=& \frac{1-\mu_3}{\ve(t)+1-\mu_3} \label{eq:systemD}
\end{eqnarray}

Let $T = K/(1-\mu_3)$ and recall that $T_3\leq T$ and $\ve(T)$ is at least $\epsS/2$ by~\refeq{eq:chooseDelta}.  We'd like to show that $T=T_3$ if and only if $h^+(T) = 0$, and then solve $h^+(T) = 0$ to obtain the desired dependency between $\epsS$ and $K$. However, under the current definition of $h^+(t)$ it is trivially 0 for any $t\geq T_3$, including $t=T$.

To remedy this, let us formally extend equations
	(\ref{eq:systemA}-\ref{eq:systemD})
to $t\in [0; 2K]$. More precisely, let us forget the original definitions of functions
	$d_3(t)$, $d_4(t)$, $h_4(t)$ and $h^+(t)$,
and \emph{redefine} them by these equations; note that these functions are uniquely determined by these equations. Obviously, the new definitions coincide with the old ones for $t\in [0, T_3]$.

By~\refeq{eq:chooseDelta2K} it follows that $h^+(t)$ is strictly decreasing for all $t\in [0; 2K]$, so $t=T_3$ is its only root. Therefore $T=T_3$
if and only if $h^+(T) = 0$. By \refeq{eq:systemD},
\begin{equation}\label{eq:hplus}
	h^+(T) = 1- (1-\mu_3) T + \int_0^T d_4(t)\, dt.
\end{equation}
Suppose we keep $K$ fixed and vary $\epsS$. Then the integral in~\refeq{eq:hplus} becomes a function of $\epsS$; let us denote it by $F(\epsS)$. This function is continuous; this is intuitively clear but nevertheless requires a rigorous proof, see Appendix~\ref{app:continuity} for details.

Note that we have $h^+(T) = 0$ if and only if $F(\epsS) = K-1$. We'll show that $F(\eps_1)>K-1$ and $F(\eps_2) < K-1$, so by continuity of $F(\epsS)$ it will follow that there exists $\ve \in (\eps_1; \eps_2)$ such that $F(\ve) = K-1$.

Let us define a function
	$$g(y) = \frac{1-\mu_3}{y+1-\mu_3}$$
and observe that it is strictly decreasing since $g'(y)<0$ for all $y$. We claim that
\begin{equation}\label{eq:DFour}
	\text{$d_4(t) \in [g(\epsS); g(\epsS/2)]$ for any $t\in [0, T]$.}
\end{equation}
Indeed, since $\ve(t)\in [\epsS/2; \epsS ]$ for all $t\in [0,T]$, it is easy to see that $h_4(t)$ is strictly increasing. Therefore the function $x(t) = t + h_4(t)$ is continuous and strictly increasing, so for any $x\in [0; T]$ there exists some $t\in [0; T]$ such that $x = x(t)$. It follows that
	$$ d_4(x) = d_4(x(t)) = g(\ve(t)) \in [g(\epsS); g(\epsS/2)],$$
claim proved.

Then $F(\ve) \in [F^-(\ve);\, F^+(\ve)]$, where
$$\left\{ \begin{array}{ccccc}
F^-(\ve) &=& T\times g(\ve)   &=& K/(1+3\ve/5) \vspace{2mm} \\
F^+(\ve) &=& T\times g(\ve/2) &=& K/(1+\ve/10)
\end{array}\right.$$

It follows that
$$\left\{ \begin{array}{ccccccc}
F \left(\frac{20}{K}\right) &\leq& F^+ \left(\frac{20}{K}\right)
	&=& \frac{K^2}{K+2} &<& K-1 \vspace{0.5mm} \\
F \left(\frac{5}{3K}\right) &\geq& F^- \left(\frac{5}{3K}\right)
	&=& \frac{K^2}{K+1} &>& K-1
\end{array}\right.$$


Therefore by continuity of $F(\ve)$ for any $K\geq 20$ there exists some
	$\ve_0\in(5/3K; 20/K)$
such that $F(\ve_0) = K-1$. Let us set $\epsS = \ve_0$.

\OMIT{
\subsection{Basic gadget: proof of Claim~\ref{cl:mu2}}
\label{sec:timeUnit}

Recall that we assume that $K\geq 20$ and borrow $\ve$ from Claim~\ref{cl:epsk}. This determines all other parameters $(\delta,\mu_3, K)$ except $\mu_2$. We'll vary $\mu_2$ in the interval $[0; \mu_3]$. Then $h_2(T_1)$, the maximal height of server $Q_2$,
 becomes a function of $\mu_2$; we denote this function by $H_2(\mu_2)$.

For any time $t\in [0; T_1]$, session $H$ fluid enters server $Q_2$ at rate 1, and leaves it at rate $1-\mu_2$, so
	$$H_2(\mu_2) := h_2(T_1) = T_1 \times \mu_2/(1-\mu_2).$$
Note that time $T_1$ is independent of $\mu_2$. Indeed, for any time $t\in [0; T_1]$ we have
\begin{equation}\label{eq:h1}
	h_1(t)	= 1+ \int_0^t (d_4(t) - 1) \, dt.
\end{equation}
Since $d_4(t)$ is determined by the system
	(\ref{eq:systemA}-\ref{eq:systemD}),
it is independent of $\mu_2$; therefore by~\refeq{eq:h1} so are height $h_1(t)$ and time $T_1$.

We claim that $1/\ve \leq T_1 \leq 3/\ve$. Indeed, at any time $t\in [0; T_1]$, server $Q_1$ drains at rate $1-d_4(t)$, which by \refeq{eq:DFour} lies between
	$\frac{\ve}{2}/(1+\frac{\ve}{2}-\mu_3)$
and
	$\ve/(1+\ve-\mu_3)$.
Hence the time $T_1$ at which server $Q_1$ drains satisfies,
$$ \frac{1}{\ve} \leq \frac{1+\ve-\mu_3}{\ve} \leq T_1 \leq
	\frac{1+\frac{\ve}{2}-\mu_3}{\frac{\ve}{2}} \leq \frac{3}{\ve},
$$
claim proved. It follows that
	$$\mu_2/\ve \leq H_2(\mu_2) \leq 6\mu_2/\ve,$$
so, in particular, $H_2(\ve/60) \leq 1/10$, and $H_2(\ve/5) \geq 1/5$.
By continuity of $H_2(\mu_2)$ there exists a value
	$\mu_2\in [\ve/60; \ve/5]$
such that
	$H_2(\mu_2) = \TP/\cel{5\TP}$.
} 

\section{Full construction}\label{sec:fullConstruction}

\newcommand{\Ncols}{\ensuremath{N_{\mathrm{cols}}}} 
\newcommand{\Nrows}{\ensuremath{N_{\mathrm{rows}}}} 
\newcommand{\Nsup}{\ensuremath{N_{\mathrm{sup}}}} 
\newcommand{\Nvert}{\ensuremath{N_{\mathrm{vert}}}}
\newcommand{\Gijk}{\ensuremath{\G_{(i,j,k)}}}

\newcommand{\epsH}{\ensuremath{\ve_{\mathrm{h}}}}
\newcommand{\epsV}{\ensuremath{\ve_{\mathrm{v}}}}
\newcommand{\deltaH}{\ensuremath{\delta_{\mathrm{h}}}}
\newcommand{\deltaV}{\ensuremath{\delta_{\mathrm{v}}}}

\newcommand{\muD}{\ensuremath{\mu_{\mathrm{D}}}} 

\newcommand{\LoadedCols}{\ensuremath{\mathcal{C}_0}}
\newcommand{\Qrepl}{\ensuremath{Q_{\mathrm{repl}}}}
\newcommand{\ConnProfile}{\ensuremath{\mathcal{C}_{\mathrm{prof}}}}
\newcommand{\IndHyp}{\ensuremath{\mathbf{H}}}

In this section we'll define the full construction and prove that it has the  periodicity claimed in Theorem~\ref{thm:main}. We'll use the basic gadget from Section~\ref{sec:gadget} as the main building block. Recall that this gadget has several tunable parameters (see Table~\ref{tab:paramsGadget}). All gadgets we use will be identical, i.e. will have the same parameters. Specifically, we set $K = 40$ and use Lemma~\ref{lm:phase} to get parameters
	$(\epsS,\deltaS,\mu_2,\mu_3, K')$
such that $\epsS \leq 1/2$, and the corresponding times
	$(T_1, T_2, T_3, \TP)$.

\subsection{Full construction: the layout}
\label{sec:layout}

As we mentioned in Section~\ref{sec:gadgetConstruction}, we organize gadgets in rows, so that the same horizontal session $H$ goes consecutively through all gadgets in a given row. Formally, a sequence of basic gadgets is merged into a \emph{row of gadgets} as follows.

Recall that each basic gadget $\G$ contains four servers $Q_i=Q_i(\G)$, and the horizontal session path that we denote $P(\G)$; recall that this path includes cycles and loops. Let
	$\G_0, \G_1, \ldots, \G_n$
be a sequence of gadgets. Firstly, two new nodes $s$, $t$ are introduced; they designated as the source and the sink of the horizontal session $H$, respectively. Secondly, for each $i$ we identify server $Q_4(\G_i)$ from gadget $\G_i$ with server $Q_1(\G_{i+1})$ from gadget $\G_{i+1}$. Thirdly, we define the session path for $H$ as
	$$(s, q) \oplus P(\G_0) \oplus \ldots \oplus P(\G_n)
		\oplus (q',t),$$
where $q = Q_1(\G_0)$ and $q'= Q_4(\G_n)$ are two servers, and $P\oplus P'$ denotes the concatenation of two paths $P$ and $P'$ such that the last node of $P$ is the first node of $P'$. In particular, session $H$ goes through each server $Q = Q_4(\G_i)$
for the total of $2K+K'$ times: first $K$ times in the big loop of gadget $\G_i$, then $K'$ times in the small loop of the same gadget, and then another $K$ times in the big loop of gadget $\G_{i+1}$.

Several consecutive rows will form a \emph{super-row}. We will have \Nsup\ super-rows, of \Nrows\ each; each row will consist of \Ncols\ gadgets. Here \Nsup, \Nrows\ and \Ncols\ are parameters that we will tune later. Let us number rows and super-rows starting from 0, in the downward direction. In a given row, let the horizontal session flow from left to right; let us number gadgets from left to right, starting from 0. Let \Gijk\ be the $k$-th gadget in the $j$-th row of super-row $i$. The \Nrows\ gadgets \Gijk, $j\in [\Nrows]$ will form a \emph{column} $C_{ik}$. So each super-row $i$ consists of \Ncols\ columns $C_{ik}$, $k\in[\Ncols]$.

Recall that we have a distinct horizontal session for each row. All horizontal sessions have the same flow control parameters: we let \epsH\ be the maximal flow rate, and \deltaH\ be the damping parameter. In addition to horizontal sessions and simple sessions from Section~\ref{sec:gadget}, we'll have a new type of session: \emph{vertical sessions}. All such sessions have the same flow control parameters: we let \epsV\ be the maximal flow rate, and \deltaV\ be the damping parameter.

Each server $Q_1$ in the very first gadget of each row is called a \emph{replenishing server}. Each vertical session $V$ goes through exactly one such server, call it $Q_V$. It will be the case that exactly \Nvert\ vertical sessions go through each replenishing server, periodically flooding it with vertical fluid, causing build-up (replenishing) of the blocked horizontal fluid. See Section~\ref{sec:dynamics} for more details.

For each replenishing server $Q$ there is a distinct new server with service rate
	$\muD\geq \Nvert \,\epsV$
that we call the \emph{decelerating server}. For each vertical session $V$ such that $Q = V_Q$ this is the last server that $V$ goes through before it reaches its sink. In case there is a build-up of vertical fluid, this server acts as a bottleneck, making sure that the build-up persists (and thus session $V$ stays blocked) for a sufficiently long time.

Each vertical session $V$ is associated with some column $C_{ik}$. Session $V$ injected into server $Q_2$ of the first gadget of this column, and goes consecutively through servers $Q_2$ of all gadgets in the column. Then $V$ goes through the corresponding replenishing server $Q_V$, then it goes through the corresponding decelerating server, and then it proceeds to its sink.

At most one vertical session passes through a given column. Some columns do not contain a vertical session; those are called \emph{blank}. We need to specify which columns are blank, and which vertical sessions connect to which replenishing queues. Let us denote this information as a collection \ConnProfile\ of all $4$-tuples $(i,k,l,j)$ such that column $C_{ik}$ contains a vertical session $V$, and the corresponding replenishing server $Q_V$ lies in row $j$ of super-row $l$. Collection \ConnProfile\ describes connectivity between the super-rows; thus we call it \emph{connectivity profile}.

\begin{figure}
\centering
\includegraphics[width=3.3in]{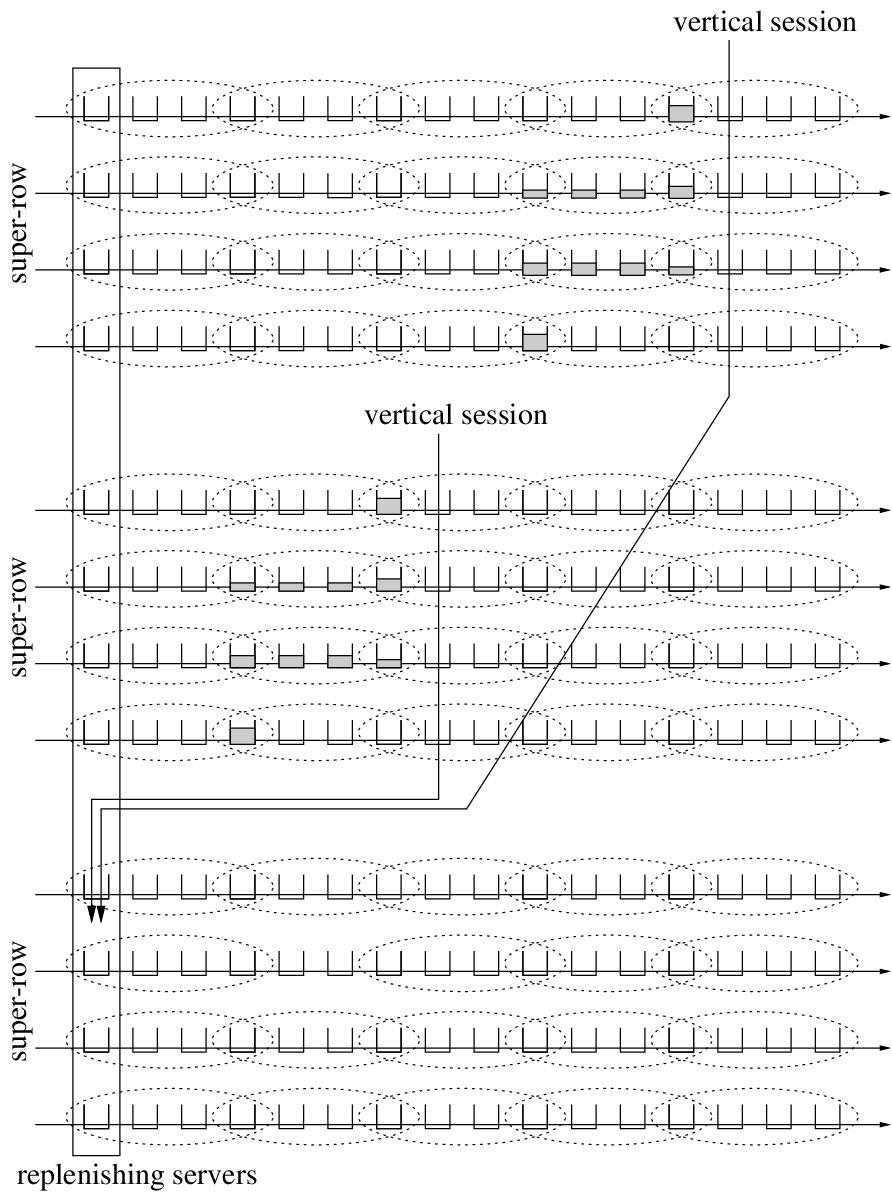}
\caption{Super-rows for $\Nrows=4$ and $\Nvert=2$.}
\label{fig:superRows}
\end{figure}

We review our construction in Figure~\ref{fig:superRows}.
We will specify the connectivity profile in Section~\ref{sec:dynamics}. We also need to choose the values for various parameters, see Table~\ref{tab:paramsFull} for a list. We do it later in this section.

\begin{table}
\renewcommand{\arraystretch}{1.3}
\begin{tabular}{l|l}
Parameter & Description   \\ \hline
$(\epsV,\deltaV)$	& flow control parameters for the vertical sessions  \\
$(\epsH,\deltaH)$	& flow control parameters for the horizontal sessions \\
\muD		& service rate for the decelerating servers, $\mu\geq \epsV$ \\
\Nvert		& \#vertical sessions in each replenishing server \\
\Ncols		& \#columns \\
\Nrows		& \#rows in a super-row	\\
\Nsup		& \#super-rows \\ \hline
$\Deltat$	& time shift between two consecutive rows in a super-row \\
$\DeltaT$	& time shift between two consecutive super-rows\\
$\DeltaT^*$	& duration of a row replenishing process
\end{tabular}

\caption{Tunable parameters in the full construction}
\label{tab:paramsFull}
\end{table}

\subsection{Full construction: the max-stable flow constraints}
\label{sec:max-stable}

We say that at a given time a session $S$ is \emph{max-stable} if it is stable and is injecting at its maximal injection rate. Let us say that a super-row $\mathcal{R}$ is in a \emph{max-stable} state if all  horizontal sessions in $\mathcal{R}$ are in a max-stable state, and all vertical sessions in the columns of $\mathcal{R}$ are in a max-stable state. To ensure that such state is possible, we need to satisfy the \emph{max-stable flow constraints}: for each server $Q$, the sum of the maximal injection rates of all sessions served by $Q$ is no more than this server's service rate. Let us list these constraints:
$$\left\{ \begin{array}{ll}
K\epsH + \epsV \Nvert \leq 1   	& \text{(at each replenishing server $Q_1$)} \\
(2K+K')\,\epsH + \epsS \leq 1 	& \text{(at each non-replenishing $Q_1$)} \\
K\epsH + \epsV \leq 1-\mu_2,	& \text{(at each server $Q_2$)} \\
K\epsH \leq 1-\mu_3.		& \text{(at each server $Q_3$)} \\
\Nvert\, \epsV \leq \muD	& \text{(at each decelerating server)}
\end{array}\right.$$
To satisfy these constraints, let us set
$$\left\{ \begin{array}{lcl}
\Nvert 	&=& \cel{1/(\mu_3-\mu_2)} \\
\epsH^* &=& 1/(8K+4K')\\
\epsV^* &=& 1/4 \Nvert
\end{array}\right.$$
and let us enforce that
\begin{equation}\label{eq:max-stable}
	\text{$\epsH\leq\epsH^*$ and $\epsV\leq\epsV^*$ and $\muD\geq 1/4$.}
\end{equation}
Since we chose \epsS, the maximal flow rate for a simple session, to be at most $1/2$,~\refeq{eq:max-stable} implies all max-stable flow constraints.

\subsection{Full construction: the replenishing process}
\label{sec:replenish}

\newcommand{\Row}{\ensuremath{\mathbf{P}}}
\newcommand{\RowEmpty}{\ensuremath{\Row_{\mathrm{empty}}}}
\newcommand{\SRow}{\ensuremath{\mathbf{S}}}
\newcommand{\SRowEmpty}{\ensuremath{\SRow_{\mathrm{empty}}}}

\newcommand{\STATEofNet}{\ensuremath{\text{\bf ST}}}
\newcommand{\BasicProcess}{\ensuremath{\text{\bf B}}}
\newcommand{\InitProc}{\ensuremath{\mathbf{I}}} 
\newcommand{\ReplProc}{\ensuremath{\mathbf{R}}} 

In this subsection we will consider the dynamics of our construction. We will build on the dynamics of the basic time phase described in Section~\ref{sec:gadget}. In particular, in a given row this process will happen in each gadget, one gadget at a time, consecutively from left to right. Different rows will follow the same process, shifted in time by a certain amount.

Let us introduce some machinery needed to describe and reason about such processes. Let us start by defining formally what is a \emph{state} in our construction. The state of session $S$ at a given time $t$ is a record $ \langle Q,t,\, \ldots \rangle$ recording the injection rate of $S$ at time $t$, whether at time $t$ the session is blocked or happy or stable, and (if it has a positive injection rate and it is not stable) the last time it switched between these three modes. The state of server $Q$ at a given time $t$ is a set of records $\langle Q,t,S,\,\ldots \rangle$ which for each session $S$ coming through server $Q$ describes the session $S$ fluid buffered at $Q$ at time $t$.

Let $\mathcal{N}$ be the set of all servers and sessions. For each subset $\mathcal{N'}\subset \mathcal{N}$ the state of $\mathcal{N'}$ is the union of states of all elements of $\mathcal{N'}$. Note that the state of $\mathcal{N}$ at time $t_0$ determines the state of $\mathcal{N}$ at any future time $t>t_0$. We will also argue about the dependencies between the states of different rows and super-rows in our construction. Finally, let us define the \emph{process} in $\mathcal{N'}\subset \mathcal{N}$ in time interval $I$ as the union of states of $\mathcal{N'}$ for all times $t\in I$.

Now we can introduce a more concrete definition of the basic time phase. In a gadget, the \emph{$\BasicProcess(t)$-process} is the basic time phase which starts at time $t$ and ends at time $t+\TP$. We will ensure that each gadget $\G_{(i,j,k)}$ undergoes the $\BasicProcess(t)$-process for each
\begin{equation}\label{eq:Gijk}
	t = (n \Nsup +i)\,\DeltaT + j\Deltat + k\TP,\; n\in\N,
\end{equation}
where parameters \DeltaT\ and \Deltat\ are the same for all gadgets. Each gadget will undergo exactly one $B(t)$-process every $\Nsup\,\DeltaT$ time units. The rest of the time it will be ``idle''; we'll make it more precise later. In particular, all gadgets in the same row will follow the same process, but the processes in two consecutive gadgets are shifted by time \TP. Similarly, all rows in the same super-row will follow the same process, but the processes in two consecutive rows are shifted by time \Deltat; all super-rows will follow the same process, but the processes in two consecutive super-row are shifted by time \DeltaT.

Recall that we fixed the value for \TP\ when we invoked Lemma~\ref{lm:phase}. We want to define \Deltat\ such that $0<\Deltat<T_2$ and the ratio $\TP/\Deltat$ is an odd integer; we will use these properties in the proof of Lemma~\ref{lm:connProfile} to enable equation~\refeq{eq:synchB}. Let us set
\begin{equation}\label{eq:Deltat}
	\text{$\Deltat := \TP/r$, where $r := 2\cel{\TP/T_2}+1$}.
\end{equation}
We will specify \DeltaT\ later in this subsection.

We introduced rows of gadgets in order to enable the following intuitive claim:
\begin{claim}
If for time $t$ some gadget in a row undergoes the $\BasicProcess(t)$-process, and the next gadget in this row is 'nice', then this next gadget undergoes the $\BasicProcess(t+\TP)$-process.
\end{claim}
Here the next gadget being 'nice' means that in this gadget, servers $Q_2$, $Q_3$ and $Q_4$ are empty of horizontal and simple sessions' fluid, and the simple session is max-stable. We will need to use this intuitive claim in the subsequent proofs about states of the full construction. For this purpoce we need to express what happens in this lemma in terms of \emph{states}, and to do that we need a somewhat more complicated formalism.

To define precisely what we mean by the state of row $\mathcal{R}$, let us think of $\mathcal{R}$ as a system that contains the horizontal session and (for each gadget) the four servers $Q_i$ and the simple session. Let us define the \emph{partial state} of row $\mathcal{R}$ as the subset of its state, namely all of the state except the records about vertical sessions coming through $\mathcal{R}$. Let us define some partial states that will be very useful.

\begin{figure}
\centering
\includegraphics[width=3.2in]{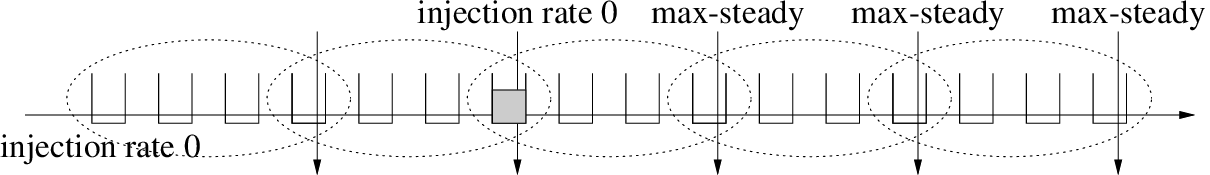}
\caption{Illustraion of a \emph{$\Row_k$-state}}
\label{fig:Pstate}
\end{figure}

In row $\mathcal{R}$, let $\G_k$ be gadget $k$ in this row. Say at a given time row $\mathcal{R}$ is in a \emph{$\Row_k$-state} where $0\leq k<\Ncols$ if
\begin{itemize}
\item all servers in row $\mathcal{R}$ are empty of horizontal or simple sessions' fluid, except server $Q_1$ in gadget $\G_k$. This server holds no simple session's fluid, and exactly one unit of horizontal fluid which is \emph{fresh} with respect to $\G_k$.

\item the injection rate into the horizontal session is 0; all simple
sessions in gadgets $\G_l$, $l\geq k$ are max-steady; if $k\geq 1$
then the simple session in gadget $\G_{k-1}$ has zero injection
rate. (See Figure~\ref{fig:Pstate}.)
\end{itemize}

\noindent Say row $\mathcal{R}$ is in a \emph{$\Row_k$-state}, $k=\Ncols$, if
\begin{itemize}
\item all servers in row $\mathcal{R}$ are empty of horizontal or simple sessions' fluid, except server $Q_4$ in gadget $\G_k$. This server holds exactly one unit of horizontal fluid which is \emph{processed} with respect to $\G_k$.

\item the injection rate into the horizontal session is 0; the simple session in gadget $\G_k$ has zero injection rate.
\end{itemize}

\begin{figure}
\centering
\includegraphics[width=3.2in]{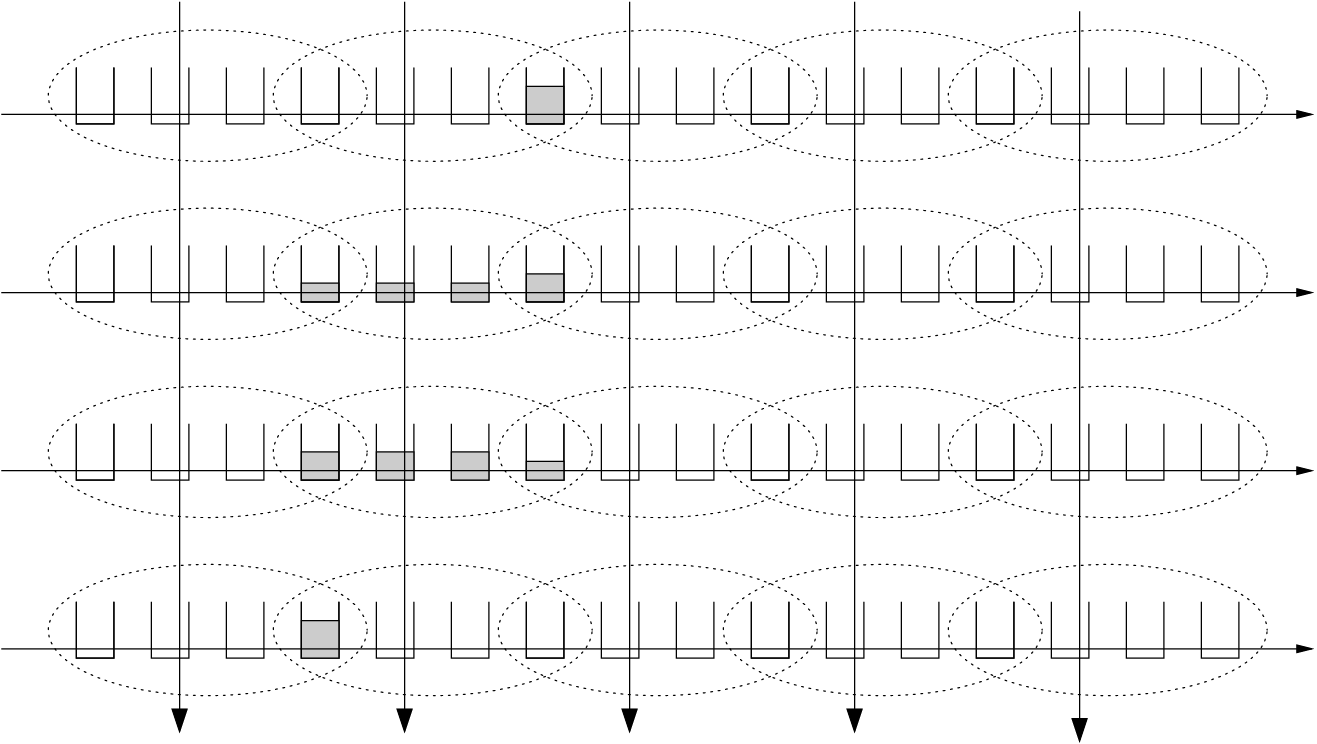}
\caption{Illustraion of an \emph{$\SRow_k$-state}}
\label{fig:Sstate}
\end{figure}

\noindent Note that $\Row_0$-state and $\Row_1$-states can be seen as a unique partial states of row $\mathcal{R}$. For $k\geq 2$ we view the $\Row_k$-state as a \emph{collection} of all partial states that satisfy the appropriate conditions.  Say that a row is \emph{empty} if all servers in this row are empty of horizontal or simple sessions' fluid; let \RowEmpty\ be the collection of all partial states of a row such that it is empty. The following lemma characterizes the partial states that occur after the row is in partial state $\Row_1$.

\begin{claim}  \label{cl:row}
Consider a row $\mathcal{R}$ in our construction. For convenience, let us denote
	$t_* = \Ncols\,\TP+1$.
\begin{itemize}
\item[(a)] Suppose at time $t$ row $\mathcal{R}$ is in a $\Row_k$-state, for some $k$ such that
	$1\leq k<\Ncols$.
Then row $\mathcal{R}$ undergoes the $\BasicProcess(t)$-process, and at time $t+\TP$ it is in a $\Row_{k+1}$-state.

\item[(b)] Suppose at time $\TP$ row $\mathcal{R}$ is in the partial state $\Row_1$. Then for each time
	$t\in [\TP, t_*]$
the partial state of this row  is fixed, call it $\Row(t)$. In particular,
	$\Row(k\,\TP)\in \Row_k$
for each $k$ such that $1\leq k \leq \Ncols$,
 and $\Row(t_*)\in\RowEmpty$.

\item[(c)] Suppose at some time $t\in [\TP,t_*]$ row $\mathcal{R}$ is in the partial state $\Row(t)$. Then for each time
	$t'\in [t, t_*]$
the partial state of this row is $\Row(t')$.
\end{itemize}
\end{claim}

In fact each row will be initialized in partial state $\Row_0$, but in this case the behavior of (the partial state of) the row depends on what happens outside this row; this is why we chose partial state $\Row_1$ as the starting point for $\Row(t)$

\begin{claim} \label{cl:SRow0}
Suppose at time $0$ row $\mathcal{R}$ is in partial state $\Row_0$, and that during time interval
	$[0;\, \TP]$
no vertical fluid enters the replenishing server of $\mathcal{R}$. Then this row undergoes the $\BasicProcess(0)$-process, and at time $\TP$ it is in partial state $\Row_1$.
\end{claim}

Let us say that a given super-row $i$ includes the corresponding rows, decelerating servers in these rows, and all vertical sessions $V_{ik}$.  Let us define the \emph{partial state} of this super-row as the partial state of all rows. For each $k\leq\Ncols$ such that
\begin{equation}\label{eq:SRow}
	k\,\TP- \Nrows\,\Deltat \geq \TP
\end{equation}
let us define a partial state $\SRow_k$ such that each row $j$ is in partial state $\Row(k\,\TP-j\,\Deltat)$. (See Figure~\ref{fig:Sstate}.)

Let us translate Claim~\ref{cl:row} into the corresponding claim into partial states of a super-row. For convenience let us introduce the following conventions:
\begin{equation}\label{eq:conventions}
\left\{ \begin{array}{lcr}
	t_* &=& \Ncols\,\TP+1  	\\
	k_0 &=& 1+\cel{\Nrows/r}\\
	t_0 &=& k_0 \TP.
\end{array}\right.
\end{equation}

\begin{claim} \label{cl:srow}
Consider a super-row $i$ in our construction.
\begin{itemize}
\item[(a)] Suppose at some time $t$ super-row $i$ is in partial state $\SRow_k$, for some $k$ such that $k_0\leq k<\Ncols$. Then at time $t+\TP$ it is in partial state $\SRow_{k+1}$.

\item[(b)] Suppose at time $k_0\,\TP$ a given super-row is in partial state $\SRow_{k_0}$. Then for each time $t\in [t_0;\, t_*]$ the partial state of this super-row is fixed, call it $\SRow(t)$. Moreover, it is the case that $\SRow(k\,\TP) = \SRow_k$ for all $k$ such that
	$k_0\leq k \leq \Ncols$.

\item[(c)] Suppose at some time $t\in [t_0;\, t_*]$ a given super-row $i$ is in partial state $\SRow(t)$. Then for each time $t'\in [t, t_*]$ this super-row is in partial state $\SRow(t')$.
\end{itemize}
\end{claim}

Let us say that a \emph{partial process} in a super-row in time interval $I$ is a union of all partial states in $I$. The above claim defines the  partial process
\begin{equation}\label{eq:partialProcess}
	\{ \SRow(t):\, t\in[t_0;\, t_*] \}
\end{equation}
For each non-blank column $C_{ik}$, let $V_{ik}$ be the vertical session that goes through $C_{ik}$. For each such session, we would like to understand how it interacts with partial process~\refeq{eq:partialProcess} in super-row $i$. Intuitively, for each row $\mathcal{R}$ of this super-row all interesting interaction with session $V_{ik}$ happens when row $\mathcal{R}$ is between partial states $\Row_k$ and $\Row_{k+1}$. For technical convenience we want partial process~\refeq{eq:partialProcess} to be well-defined whenever any row is between these two partial states. Therefore we will enforce that for any vertical session $V_{ik}$ we have
\begin{equation}\label{eq:bound-k}
	k_0 \leq k < \Ncols - k_0
\end{equation}

\OMIT{ 
To make some of the forth-coming proofs somewhat simpler, the first inequality is somewhat stronger than what we need here to satisfy~\refeq{eq:SRow} (namely, than $k\geq k_0$).
} 

In partial process~\refeq{eq:partialProcess} all horizontal fluid eventually drains to its sinks. As we want our construction to exhibit oscillations, we need a way to get back to partial state $\SRow_{k_0}$. Let us consider this problem for each row separately: we start with an empty row, and we want to get back to partial state $\Row_1$; most crucially, we need to replenish the horizontal fluid in server $Q_1$ in column 0.

Before we can state our basic result on replenishing, we need to build some machinery. Say row $\mathcal{R}$ is \emph{max-stable} if the horizontal session and all simple sessions are max-stable, and all servers contain no horizontal or vertical fluid. We would like to define a process that turns row $\mathcal{R}$ from a max-stable state to partial state $\Row_1$; we will call it a \emph{row replenishing process}. Apart from partial state $\Row_1$, we need to ensure that row $\mathcal{R}$ behaves 'nicely' towards vertical sessions that come through $\mathcal{R}$.  Specifically, say row $\mathcal{R}$ is \emph{open} if for each non-blank column in this row,
server $Q_2$ (in the gadget in this column) has at least $\epsV$ units of capacity for the vertical session that flows through this column.

Let us formulate the initial conditions for our replenishing
process. Recall that each vertical session $V$ goes through some
replenishing server $Q_V$. For a given replenishing server $Q$ in row
$\mathcal{R}$, let us define the inverse mapping $V_Q$ as the set of
all vertical sessions $V$ such that $Q = Q_V$. Say our construction is
\emph{$Q$-synchronized} if at time $t$ the following conditions hold:
\begin{equation}\label{eq:defnSynch}
\begin{array}{l}
	\text{Row $\mathcal{R}$ is max-stable; for each session $V_{ik}\in V_Q$}\\
	\;\;\;\text{$V_{ik}$ is max-stable and satisfies~\refeq{eq:bound-k}, }\\
	\;\;\;\text{and super-row $i$ is in partial state $\SRow_k$.}
\end{array}\end{equation}

The main result of this subsection is (essentially) that if our construction is $Q$-synchronized, then a certain time later row $\mathcal{R}$ is in partial state $\Row_1$.

\begin{lemma} \label{lm:repl}
There exist parameters
	$(\epsH,\deltaH,\epsV,\deltaV,\muD)$
and positive integers $(L_1, L_2)$ such that the max-stable flow constraints~\refeq{eq:max-stable} hold, and for
	$\Nrows = r\,L_1$
and any choice of
	$(\DeltaT,\Ncols,\Nsup,\ConnProfile)$
we have the following property:
\begin{itemize}
\item[(*)] Let $Q$ be a replenishing server such that $|V_Q| = \Nvert$. Suppose our construction is $Q$-synchronized at time $t$. Then at time
	$t+\DeltaT^*$ server $Q$'s row is in partial state $\Row_1$, and each session $V_{ik}\in V_Q$ is happy. Moreover, between time $t$ and $t+\DeltaT^*$ this row is open. Here
\begin{equation}\label{eq:DeltaTprop}
	\DeltaT^* = (L_1+L_2+1)\,\TP - \Deltat.
\end{equation}
\end{itemize}
\end{lemma}

We defer the proof to Section~\ref{sec:lm-repl}. We would like to make several remarks. First, note that property (*) is conditional: in this lemma we do not prove that our construction ever becomes $Q$-synchronized. Second, we achieved this property by tuning parameters
	$(\epsH,\deltaH,\epsV,\deltaV,\muD,\Nrows)$.
Note that we did not impose any constraints on parameters
	$(\DeltaT,\Nsup, \Ncols, \ConnProfile)$.
Third, we had a lot of flexibility in choosing the separation time $\DeltaT^*$. We used this flexibility to enforce~\refeq{eq:DeltaTprop}, which we will use in order to make our construction $Q$-synchronized at the appropriate times. Specifically, we will use~\refeq{eq:DeltaTprop} in the proof of Lemma~\ref{lm:connProfile} in order to pass from~\refeq{eq:synch} to \refeq{eq:synchA}.

Let us conclude this subsection by attaching appropriate names to the process described in Lemma~\ref{lm:repl}.
Let us say that a \emph{$(Q,t)$-replenishing process} is (any) process that happens in our construction between times $t$ and $t+\DeltaT^*$ if at time $t$ it becomes $Q$-synchronized. By Lemma~\ref{lm:repl} during a $(Q,t)$-replenishing process
the corresponding row goes from being empty at time $t$ to partial state $\Row_1$ at time $t+\DeltaT^*$. Informally, we say that this row gets \emph{replenished}.

Let us define the analogous replenishing process for a given super-row $i$. Let $Q^{(j)}$ be the replenishing server in row $j$ of this super-row, so that all rows experience the same replenishing process, appropriately time-shifted. Let us say that a \emph{$(i,t)$-replenishing process} is (any) process in our construction that starts at time $t$ and proceeds for time
	$(\Nrows-1)\Deltat+\DeltaT^*$,
such that for each row $j$ in super-row $i$, a
	$(Q^{(j)},\, t+j\,\Deltat)$-replenishing process happens.

\subsection{Full construction: proof of Lemma~\ref{lm:repl}}
\label{sec:lm-repl}

\newcommand{\TStop}{\ensuremath{T_{\text{stop}}}}
\newcommand{\Bv}{\ensuremath{B_{\text{v}}}}
\newcommand{\Bh}{\ensuremath{B_{\text{h}}}}

Consider a session with maximal flow rate $\ve$ and damping parameter $\delta$. Recall that $f_\delta$ is the parameterized flow decrease function (see Section~\ref{sec:flowControl}). Suppose at time $0$ the session is injecting at rate $\ve$, then becomes blocked and stays blocked. Then by axiom~\refAS{ass:decrease} the injection rate starts decreasing until at some finite time
	$\TStop = \TStop(\ve,\delta)$
it becomes $0$. Let
	$$ A(\ve, \delta, t) := \int_0^t f_\delta(\ve, t) \, dt $$
be the amount of fluid that is injected by time $t$.  Let
	$$ B(\ve, \delta) := \int_0^\infty f_\delta(\ve, t) \, dt  = A(\ve,\delta, \TStop)$$
be the amount of fluid that is injected by the time the flow stops. Let
	$\Bv = B(\epsV, \deltaV)$ and $\Bh = B(\epsH, \deltaH)$
be these amounts for vertical and horizontal sessions, respectively.

Let us fix a choice of $(\Ncols,\Nsup,\ConnProfile)$, suppose $Q$ is a replenishing server such that with respect to this choice $|V_Q| = \Nvert$, and assume that our construction is $Q$-synchronized at time $t$. Let us focus on some vertical session
	$V = V_{ik} \in V_Q$.
The path of this session goes through \Nrows\ consecutive gadgets in column $C_{ik}$. Let
	$\G_j = \G_{(i,j,k)}$
be the $j$-th such gadget, and let $Q^{(j)}$ be the server $Q_2$ from $\G_j$. Say such server is \emph{blocked} if it stores a positive amount of horizontal fluid. Then, since at servers $Q^{(j)}$ the horizontal fluid has priority over the vertical fluid, session $V$ is blocked if at least one server $Q^{(j)}$ is blocked.

Recall that at time $t$ super-row $i$ is in partial state $\SRow_k$. Let us consider what happens with session $V$ after time $t$. Until time $t$ session $V$ is in the max-stable state. Then each server $Q^{(j)}$ is blocked  starting from time $t+j\,\Deltat$ for time $T_2$, so session $V$ is blocked starting from time $t$ for time
\begin{equation}\label{eq:tstar}
	t^* := (\Nrows-1)\, \Deltat+ T_2.
\end{equation}
We will make sure that by this time the injection rate into session $V$ goes from \epsV\ to $0$:
\begin{equation}\label{eq:TStopV}
	\TStop(\epsV, \deltaV) \leq t^*.
\end{equation}
Then \Bv\ units of session $V$ fluid is accumulated in server $Q$ at time  $t+t^*$. From this time on, all servers $Q^{(i)}$ become unblocked; in each of them, horizontal fluid eats up at most $1-\mu_3$ units of bandwidth (this is by Lemma~\ref{lm:phase}), so at least $\mu_3-\mu_2$ units of bandwidth are available to session $V$. Therefore, starting from time $t+t^*$ session $V$ fluid drains down into server $Q$ at rate at least $\mu_3-\mu_2$, until at some time $t_2$ all \Bv\ units of fluid are gone.

Since this happens for all \Nvert\ vertical sessions $V\in V_Q$, between times $t+t^*$ and $t_2$ the total incoming rate of vertical sessions into server $Q$ will be at least
	$ \Nvert (\mu_3-\mu_2)$,
which is at least $1$ by definition of \Nvert. Therefore at time $t+t^*$ vertical fluid will immediately start building up in server $Q$. Since at $Q$ the vertical fluid has priority over the horizontal fluid, this build-up will persist exactly for time
	$\Nvert\, \Bv$.
During all this time vertical fluid exits server $Q$ and enters the corresponding decelerating server $D_Q$ at rate $1$. We will choose the service rate $\muD$ for $D_Q$ so that $\muD < 1$. Therefore after leaving server $Q$, vertical fluid builds up in server $D_V$ and, moreover, this build-up drains down to the sink only \emph{after} all vertical session fluid drains down from $Q$. Therefore a total of $\Nvert\,\Bv$ units of vertical fluid drains down server $D_V$, at a constant rate $\muD$, taking time $\Nvert\,\Bv/\muD$. During this time interval, the injection rate into each session $V\in V_Q$ is $0$.  We will ensure that this time interval is sufficiently long, namely that
\begin{equation}\label{eq:muD}
	\Nvert\,\Bv/\muD = \Nvert\, \Bv + \TP.
\end{equation}	
Note that if~\refeq{eq:muD} holds then $\muD < 1$.

Let $H$ be the horizontal session coming through server $Q$. Before
time $t+t^*$ session $H$ is in the max-stable state. Recall that
server $Q$ holds a non-zero amount of vertical fluid starting from
time $t+t^*$, for time $\Nvert\, \Bv$. During this time interval,
session $H$ is blocked. We will tune the parameters so that during
this time interval the injection rate of session $H$ goes from \epsH\ down to $0$:
\begin{equation}\label{eq:TStopH}
	\TStop(\epsH, \deltaH) \leq \Nvert\, \Bv,
\end{equation}	
and meanwhile exactly $1$ unit of session $H$ fluid accumulates in server $Q$:
\begin{equation}\label{eq:Bh}
	\Bh = 1.
\end{equation}	

If equations~\refeqs{eq:muD}{eq:Bh} hold, then at time
	$t+t^*+\Nvert\, \Bv$
the row $\mathcal{R}$ containing server $Q$ in partial state $\Row_0$. By Claim~\ref{cl:SRow0}  this row is in partial state $\Row_1$ time $\TP$ later, i.e. at time $t+\DeltaT^*$ where
\begin{equation}\label{eq:DeltaT}
	\DeltaT^* = t^*+\Nvert\, \Bv +\TP.
\end{equation}
Moreover, by~\refeq{eq:muD} at this time the decelerating server of
row $\mathcal{R}$ is empty, and therefore all vertical sessions $V\in
V_Q$ are happy. Row $\mathcal{R}$ is open between time $t$ and
$t+\DeltaT^*$ since there was no build-up of horizontal fluid at time $t$, and the horizontal session stayed blocked between time $t$ and $t+\DeltaT^*$.

To complete the proof of Lemma~\ref{lm:repl}, it remains to match~\refeq{eq:DeltaT} with the definition~\refeq{eq:DeltaTprop} of $\DeltaT^*$.

Note that by definition~\refeq{eq:tstar} of $t^*$ and since
$$ \Nrows\, \Deltat = L_1 \cel{\TP/T_2}\, \Deltat = L_1\, \TP,
$$
we can write $t^*$ as a function of $L_1$:
$$ t^* =  L_1\,\TP + T_2-\Deltat.
$$
In particular, by~\refeq{eq:DeltaT} equation~\refeq{eq:DeltaTprop} becomes equivalent to
\begin{equation}\label{eq:Bv}
	\Nvert\, \Bv = L_2\, \TP - T_2.
\end{equation}

Now to prove Lemma~\ref{lm:repl} it remains to choose parameters
	$(\epsH,\deltaH,\epsV,\deltaV,\muD)$
and integers $(L_1, L_2)$ so that equations~\refeq{eq:max-stable},~\refeqs{eq:TStopV}{eq:Bh} and~\refeq{eq:Bv} hold. We need to be careful in order to avoid circular dependencies between the parameters. This is how we overcome this hurdle:

\begin{itemize}

\item[1.] choose $(\epsH, \deltaH)$ so that $\epsH\leq\epsH^*$ and $\Bh = 1$.
\item[2.] define
\begin{eqnarray*}
	L_2 &:=& 1+\cel{(\TStop(\epsH, \deltaH)+T_2)/\TP} \\
	T_0 &:=& (L_2\,\TP - T_2)/\Nvert.
\end{eqnarray*}
\item[3.] choose $(\epsV, \deltaV)$ so that $\epsV\leq \epsV^*$ and $\Bv = T_0$.
\item[4.] choose $L_1$ large enough so that~\refeq{eq:TStopV} holds.
\item[5.] choose $\muD$ so that~\refeq{eq:muD} holds.
\end{itemize}

Note that at the third step we ensure that~\refeq{eq:TStopH} and~\refeq{eq:Bv} hold. At the fifth step, by~\refeq{eq:Bv} we have $\Nvert\, \Bv>T_P$, so $\muD>1/2$ as required. It remains to show that we can indeed do the first and the third step. We establish this via the following claim:

\begin{claim}\label{cl:params}
If axioms~\refASS{ass:increase}{ass:continuity} hold, then for any given positive $(\ve^*, b)$ there exist flow control parameters $(\ve,\delta)$ such that
	$\ve\leq \ve^*$ and $B(\ve,\delta) = b$.
\end{claim}

\begin{myproof}
Choose parameter $\delta$ so that $B(\ve^*,\delta) \geq b$, i.e. at least $b$ units of fluid are injected if the session becomes blocked when it is transmitting at rate $\ve^*$; such $\delta$ exists by axiom~\refAS{ass:power}. Now that $\delta$ is fixed, let us tune $\ve$ in the interval $(0,\ve^*]$ so that \emph{exactly} $b$ units of fluid are injected. We do it rigorously as follows.

Let us write $\TStop(\ve) = \TStop(\ve,\delta)$; by Axiom~\refAS{ass:monotonicity} this is an increasing function of $\ve$. The function
	$$ A(\ve) := A(\ve,\,\delta,\, \TStop(\ve^*))$$
is continuous by axiom~\refAS{ass:continuity}, see Fact~\ref{fact:integral} for rigor. Since $A(0) = 0$ and $A(\ve^*) \geq b$, we can choose
	$\ve\leq \ve^*$ such that $A(\ve) = b$.
Since $\TStop(\ve)\leq \TStop(\ve^*)$, it follows that
	$B(\ve,\delta)= b$, too.
\end{myproof}

This completes the proof of Lemma~\ref{lm:repl}.

\subsection{Full construction: the connectivity profile}
\label{sec:ConnProfile}

In this subsection we will choose the connectivity profile \ConnProfile. We will assume that
the parameters are chosen as per Lemma~\ref{lm:repl}, and that
	$\DeltaT = 2\DeltaT^*$.

Recall that \ConnProfile\ is the collection of all $4$-tuples $(i,k,l,j)$ such that vertical session $V_{ik}$ goes through the replenishing server which lies in row $j$ of super-row $l$. Let us impose some simple conditions on the connectivity profile.

\begin{definition}\label{def:wellFormed}
Let us say the connectivity profile is \emph{well-formed} if it has all of the following properties:
\begin{itemize}
\item[(a)] for each pair $(i,k)$ there is at most one $(i,k,l,j) \in \ConnProfile$.
\item[(b)] for each tuple $(i,k,l,j) \in \ConnProfile$ we have $k\geq k_0+1$.
\item[(c)] for each replenishing server $Q$ we have $|V_Q| = \Nvert$.
\item[(d)] \ConnProfile\ is not changed by a circular permutation of super-rows:
	$(i,k,l,j) \in \ConnProfile$ if and only if
	$(0,k,l^*,j) \in \ConnProfile$,
for $l^* = (l-j) \pmod{\Nsup}$.
\end{itemize}
\end{definition}

Let us reverse-engineer some other useful properties of \ConnProfile\ from the requirement that for each replenishing queue $Q$, our construction needs to be $Q$-synchronized at the appropriate time. Let us go back to~\refeq{eq:Gijk}; this is when we would like a given gadget to undergo a $\BasicProcess(t)$ process. It follows that for each column $C_{ik}$ we would like the corresponding super-row $i$ to be in partial state $\SRow_k$ at time
	$$ \tau_{ik} := i\,\DeltaT+k\,\TP.$$
Similarly, for each replenishing server $Q$ in row $j$ of super-row $i$ we would like this row to be in partial state $\Row_1$ at time
	$$ \tau_Q := i\,\DeltaT+j\Deltat+\TP.$$
It follows that we would like our construction to be $Q$ synchronized at time $\tau_Q-\DeltaT^*$. In particular, for each vertical session $V_{ik}\in V_Q$ we want
\begin{equation}\label{eq:tauQ}
	\tau_{ik} \equiv \tau_Q-\DeltaT^* \pmod{\Nsup\, \DeltaT}.
\end{equation}
Say the \ConnProfile\ is \emph{synchronized} if~\refeq{eq:tauQ} holds for each replenishing queue $Q$ and each vertical session $V_{ik}\in V_Q$.

Note that a well-formed connectivity profile \ConnProfile\ is completely determined by $\Nsup$
and the set $\ConnProfile^*$ of triples $(k,l,j)$ such that
	$(0,k,l,j) \in \ConnProfile$.
Let us call this set the \emph{reduced} connectivity profile.

For a given reduced connectivity profile $\ConnProfile^*$ let us define
\begin{eqnarray*}
	\Nsup(\ConnProfile^*)  &:=& \max\{l\in\N:\, (k,l,j)\in\ConnProfile^* \} \\
	\Ncols(\ConnProfile^*) &:=& \max\{k\in\N:\, (k,l,j)\in \ConnProfile^* \}
\end{eqnarray*}
to be, respectively, is the maximal super-row that super-row $0$ is connected to, and the maximal non-blank column. Then we can choose any
	$\Nsup>\Nsup(\ConnProfile^*)$
and reverse-engineer the corresponding connectivity profile \ConnProfile. Note that if it is well-formed then we satisfy~\refeq{eq:bound-k} if and only if
\begin{equation}\label{eq:Ncols}
	\Ncols\geq \Ncols(\ConnProfile^*) + L_1+1.
\end{equation}

Clearly, being well-formed and being synchronized are properties of $\ConnProfile^*$. More precisely, for a given $\ConnProfile^*$ the connectivity profile is well-formed (resp. synchronized) for some $\Nsup>\Nsup(\ConnProfile^*)$ then it is well-formed (resp. synchronized) for all such \Nsup; in this case say that $\ConnProfile^*$ is well-formed (resp. synchronized). It turns out that these two properties are all we need from $\ConnProfile^*$.

\begin{lemma}\label{lm:connProfile}
If $\DeltaT = 2\DeltaT^*$, where $\DeltaT^*$ is given by~\refeq{eq:DeltaTprop}, then there exists a reduced connectivity profile $\ConnProfile^*$ which is well-formed and synchronized.
\end{lemma}

\begin{myproof}
Recall that within a given super-row, rows are numbered from $0$ to $\Nrows-1$. Let us write row numbers as $ir+j$, where $j\in[r]$. Note that $\ConnProfile^*$ is synchronized if and only if for each triple
	$(k,l,ir+j)\in \ConnProfile^*$
we have
\begin{equation}\label{eq:synch}
	  k\,\TP + \DeltaT^* = l\,\DeltaT + (ir+j)\,\Deltat +\TP.
\end{equation}
Indeed, this matches~\refeq{eq:tauQ} since $\tau_{0k}= k\,\TP$ and the right-hand side of~\refeq{eq:synch} is $\tau_Q$, where $Q$ is the replenishing server in row $ir+j$ of super-row $l$. Let
	$L=L_1+L_2$.
Then by equations~\refeq{eq:Deltat} and ~\refeq{eq:DeltaTprop} we can rewrite~\refeq{eq:synch} as follows:
\begin{equation}\label{eq:synchA}
	kr = (2l-1)Lr+ (2l+i)\,r - (2l-1-j).
\end{equation}
Equivalently, for some $\alpha\in\N$ we have
\begin{equation}\label{eq:synchB}
\left\{\begin{array}{rcl}
	2l &=& \alpha r+j+1 \\
	k &=& L(\alpha r + j )+ \alpha\, (r-1) + (i+j+1).
\end{array}\right.
\end{equation}
Note that in the above equation we want $l$ and $k$ to be integer; this is why in~\refeq{eq:Deltat} we defined $r$ to be integer. Also, note that the first equation in~\refeq{eq:synchB} cannot hold for odd $j$ if $r$ is even; this is why in~\refeq{eq:Deltat} we defined $r$ to be odd.

Let $A_j$ be the set of the first \Nvert\ integers that are at least
$3$ and have the same parity as $1+j$. Let us define $\ConnProfile^*$ as the collection of all triples $(k,l,ir+j)$ such that $\alpha\in A_j$, $i\in[L_1]$, $j\in[r]$, and $(k,l)$ are given by~\refeq{eq:synchB}.

Since $r$ is odd, such $\ConnProfile^*$ is synchronized. Let us check that $\ConnProfile^*$ is well-formed. We need to check the four properties in Definition~\ref{def:wellFormed}. Properties (b) and (d) are easy: the former holds since the $\alpha$'s are at least $3$, and the latter holds since we reverse-engineer \ConnProfile\ from $\ConnProfile^*$.

To prove property (a), let us consider equation~\refeq{eq:synch}. By~\refeq{eq:DeltaTprop} we have $\DeltaT>\Nrows\,\Deltat$, so  for a given choice of $k$ there can be at most one value $l$ such that~\refeq{eq:synch} holds, hence at most one triple $(k,l,ir+j)\in \ConnProfile^*$. Therefore property (a) holds.

It remains to check property (c), i.e. that for each replenishing server $Q$ we have
	$|V_Q| = \Nvert$.
Indeed, say server $Q$ lies in row $ir+j$ of super-row $l$; for simplicity let us assume
	$l>\Nsup(\ConnProfile^*)$.
Then the induced connectivity profile \ConnProfile\ contains a tuple $(n,k,n',ir+j)$ if and only if for some $\alpha\in A_j$ the pair $(l,k)$ satisfies~\refeq{eq:synchB} where the super-row number $l$ is
	$$l = (n'-n) \pmod{\Nsup}.$$
Claim follows since there are exactly \Nvert\ such pairs.
\end{myproof}

\subsection{Full construction: safe states}
\label{sec:partialStates}

Consider a given vertical session $V_{ik}$. Suppose super-row $i$ is in partial state $\SRow(t_0)$ at time $\tau_{i0}+t_0$. As far as we are concerned, Lemma~\ref{lm:repl} specifies the behavior of session $V_{ik}$ between times $\tau_{ik}$ and $\tau_{ik}+\DeltaT^*$. Here we investigate what happens with $V_{ik}$ during the rest of the time interval $[t_0;\, t_*]$, and also when super-row $i$ is empty.

Suppose a vertical session $V$ comes through a server $Q$. Let us say that at a given time server $Q$ is \emph{open} for session $V$ if $Q$ has at least $\epsV$ units of capacity available for $V$. Suppose a given super-row contains session $V$. Let us say that at a given time this super-row is \emph{open} for  session $V$ if each server in this super-row is open for $V$.

Say a vertical session $V$ is \emph{safe} at a given time if it enters the corresponding replenishing server $Q_V$ at rate at most \epsV.  Say a replenishing server $Q$ is \emph{safe} at a given time if it is empty of vertical fluid, and all vertical sessions $V\in V_Q$ are safe. Say a super-row is \emph{safe} if each replenishing server in this super-row is safe. Intuitively, if vertical sessions, replenishing servers and super-rows are safe then they do not cause any perturbations in our construction.

Let $\tau_{ik}' := \tau_{ik} + (\Nrows-1)\Deltat$.

\begin{lemma}\label{lm:safe}
Consider a vertical session $V_{ik}$. Suppose for some $t\in [t_0;\, t_*]$ at time $\tau_{i0}+t$ super-row $i$ is in partial state $\SRow(t)$ and session $V_{ik}$ is max-stable. Then
\begin{itemize}
\item[(a)] session $V_{ik}$ is safe during the time interval(s)
$$ 	\left[\tau_{i0} + t \,;\,  \tau_{i0} + t_* 	\right] \,\setminus\,
	\left[\tau_{ik}' \,;\, \tau_{ik}+ \DeltaT^* 	\right],$$
\item[(b)] super-row $i$ is open for $V_{ik}$ during the time interval(s)
$$ 	\left[\tau_{i0} + t \,;\,  \tau_{i0} + t_* 	\right] \,\setminus\,
	\left[\tau_{ik} \,;\, \tau_{ik}'		\right].$$
\item[(c)] if at time $\tau_{i0}+t_*$ super-row $i$ is safe and session $V_{ik}$ becomes unsafe at a later time $t'$, then super-row $i$ must have become unsafe during time interval
	$(\tau_{i0}+t_*;\, t')$.
\end{itemize}
\end{lemma}

\begin{myproof}
The proof of parts (ab) is in-lined in the proof of Lemma~\ref{lm:repl}. For part (c), note that at time $t_*$ both super-row $i$ and session $V_{ik}$ are safe, and by~\refeq{eq:bound-k} there is no build-up of horizontal fluid in columns $k'\leq k$ of super-row $i$. Session $V_{ik}$ can become unsafe only (a positive time after) some build-up of horizontal fluid appears in some row $\mathcal{R}$ in column $k$. The latter can happen only if at some earlier time, in row $\mathcal{R}$ the horizontal session is injecting at the positive rate, then becomes blocked by vertical fluid in its replenishing server $Q$. The latter can happen only after server $Q$ becomes unsafe.
\end{myproof}

Say a super-row is \emph{empty} at a given time if all rows in this super-row are empty. We will also need a version of Lemma~\ref{lm:safe}(c) where initially super-row $i$ is empty, as opposed to being in one of the partial states $\SRow(t)$.

\begin{lemma}\label{lm:safe+empty}
Consider a vertical session $V_{ik}$. Suppose at some time $t$ super-row $i$ is safe and empty, and session $V_{ik}$ is safe. If session $V_{ik}$ becomes unsafe at a later time $t'$, then super-row $i$ must have become unsafe in time interval $(t;\, t')$.
\end{lemma}

The proof follows that of Lemma~\ref{lm:safe}(c). The way we are going to apply Lemma~\ref{lm:safe}(c) and Lemma~\ref{lm:safe+empty}, it is crucial that super-row $i$ becomes unsafe strictly before time $t'$.

\subsection{Full construction: the high-level layout}
\label{sec:highLayout}

In this subsection we complete the specification of our construction. We choose parameters
	$(\epsH,\deltaH,\epsV,\deltaV,\muD,\Nrows)$ and positive integers $(L_1,L_2)$ from Lemma~\ref{lm:repl}. We set $\DeltaT = 2\,\DeltaT^*$, where $\DeltaT^*$ is defined by~\refeq{eq:Deltat}. We choose $\ConnProfile^*$ from Lemma~\ref{lm:connProfile}. We define \Ncols\ to be the smallest integer such that~\refeq{eq:Ncols} holds and
\begin{equation}\label{eq:NcolsProp}
	\text{\DeltaT\ divides $\Ncols\,\TP$.}
\end{equation}
Such \Ncols\ exists because $L_1\,\DeltaT$ is divisible by \TP\ by~\refeq{eq:DeltaTprop}.
We need~\refeq{eq:NcolsProp} for technical convenience, so that in later proofs we could consider moments of time when every super-row is either empty or in some well-defined partial state $\SRow(t)$.

We choose
	$\Nsup>\Nsup(\ConnProfile^*)$
large enough so that all sessions are in the happy state long enough to build up the injection rate to the maximal value. Specifically, we let
	$$\Nsup = \Ncols\,\TP\,/\,\DeltaT + n+1,$$
where $n$ is the smallest integer such that if a horizontal session and a vertical session and a simple session are in a happy state from time $0$ onward, then by time $n\,\DeltaT$ they will be injecting at their respective maximal rates. This completes the construction.

Let us check that $\Nsup\geq\Nsup(\ConnProfile^*)$, as required by definition of $\ConnProfile*$. Indeed, let $\Nsup(\ConnProfile^*) = l$ so that $ (k,l,\cdot)\in \ConnProfile^*$ for some $k$. Since the connectivity profile is synchronized, \refeq{eq:synch} holds. It follows that
	$$ \Nsup -1 \geq k\,\TP/\DeltaT \geq l-1, $$
claim proved.

\subsection{Full construction: the high-level dynamics}
\label{sec:dynamics}

\newcommand{\WellDriven}{well-driven}
\newcommand{\BigPeriod}{\ensuremath{T_{\textrm{big}}}}

In this subsection we put all pieces together and show that starting from a suitable initial state our construction exhibits oscillations, with period
	$\BigPeriod = \Nsup\,\DeltaT$.
To distinguish it from the 'local' period \TP, let us call \BigPeriod\ the \emph{big} period.

For $n\in N$, define $i_n = n \pmod{\Nsup}$.
Recall the conventions~\refeq{eq:conventions}.
We want our system to have the following behavior:
\begin{equation}\label{eq:behav}
\begin{array}{l}
	\text{For each $n\in\N$, at time $n\,\DeltaT+k_0\,\TP$} \\
	\;\;\;\text{super-row $i_n$ is in partial state $\SRow_{k_0}$.}
\end{array}\end{equation}

Moreover, we want super-row $i_n$ to get to partial state $\SRow_{k_0}$ as a result of an $(i,t)$-replenishing process, for the appropriately chosen time $t$.

If~\refeq{eq:behav} holds, then by Claim~\ref{cl:srow} it is the case that at each time
	$t\in n\,\DeltaT + [k_0\,\TP\, ,t_*]$
super-row $i_n$ is in state $\SRow(t)$, so that
	$\SRow(t_*)\in \SRowEmpty$.
Moreover, we want this super-row to stay empty (almost) till time
	$n\,\DeltaT + \BigPeriod$.

We want to prove oscillations via an inductive argument. Generally, for some time $t_0$ we want to define some state $\IndHyp_n$ such that at time $t_0+n\,\DeltaT$ we want our construction to be in state $\IndHyp_n$; for a given $n$ this will be our inductive hypothesis. The sequence
	$\{\IndHyp_n:\,n\in\N\}$
should have period \BigPeriod. We need to prove the inductive step from $n=m$ to $n=m+1$. Then we get oscillations if at time $t_0$ we initialize our construction in state $\IndHyp_0$.

With the above plan in mind, we have three difficulties to
overcome. First, to define any reasonable $\IndHyp_n$ we need to
define \emph{states} of super-rows, not partial states. For this we
need to specify exactly the behavior of the vertical sessions so that
this behavior is consistent across different super-rows; it is
somewhat non-trivial to specify this behavior in a brute-force
way. Second, we know how to argue about partial states $\SRow(t)$, so
in $\IndHyp_n$ we would like each super-row to be in one of these
partial states. For example, if~\refeq{eq:behav} holds for $n=0$, then
the earliest time super-row $0$ is in one of these partial states
(namely, $\IndHyp_{k_0}$) is time $k_0\,\TP$. Third, for technical
convenience in a given inductive step we want to worry only about a
single replenishing process, namely the one for super-row $i_n$. In
particular, we would like it to be the case that at time
$t_0+n\,\DeltaT$ the corresponding replenishing process for super-row
$i_n+1$ has not yet started. These considerations motivate the choice of $t_0
:= k_0\,\TP$ and $\DeltaT := 2\,\DeltaT^*$.

For a given super-row, we are going to define a process $\{\SRow^*(t):\,t\geq t_0\}$ such that state $\SRow^*(t)$ induces partial state $\SRow(t)$ for any time
	$t\in [t_0, t_*]$.
We want this process to be consistent with the behavior of our construction if~\refeq{eq:behav} holds for all $n\in N$. To achieve this, we will modify a super-row so that it becomes a (suitable) stand-alone system, and in this system we will define a process by specifying the (suitable) initial state at time $t_0$ and letting the system run. We let $\SRow^*(t)$ be the state of this system at time $t$.

Let us consider a super-row $\mathcal{S}$ as a stand-alone system that includes all vertical sessions coming through the columns of $\mathcal{S}$, and all decelerating servers corresponding to the replenishing servers located in $\mathcal{S}$. Note that system $\mathcal{S}$ does not include the vertical sessions that go to these replenishing servers. Recall that in the full construction for each vertical session $V$ in $\mathcal{S}$ there would be a decelerating server $D$ (in a different super-row) which session $V$ goes to. Essentially this server determines for how long session $V$ remains blocked. Since here we are trying to emulate the behavior of a super-row in the full construction, let us add such servers artificially. Specifically, let us assume that for each vertical session $V$ in $\mathcal{S}$ there is a distinct new \emph{quasi-decelerating} server $D_V$ with service rate $\muD/\Nvert$, such that session $V$ leaves $\mathcal{S}$, then goes to $D_V$, and then goes to its sink. Then $\mathcal{S}$ is a fully specified dynamical system.

Let us define the initial state $\SRow^*_0$ for system $\mathcal{S}$ such that it is in partial state $\SRow_{k_0}$, all replenishing servers are empty, and all vertical sessions in $\mathcal{S}$ are in the max-stable state. (It follows that in state $\SRow^*_0$ all servers in $\mathcal{S}$ are empty of vertical fluid.) Suppose we initialize system $\mathcal{S}$ at time $t_0$ in state $\SRow^*_0$, and let it run according to its control mechanism. Define $\SRow^*(t)$ to be the state of this system at time $t$. Note that at time $t_0$ system $\mathcal{S}$ is in partial state $\SRow_{k_0}$, so by Claim~\ref{cl:srow} at any time $t\in [t, t_*]$ it is in partial state $\SRow(t)$, as required.

Let us go back to our construction and ask how can we make sure that if we start a given super-row in state $\SRow^*_0$, it will indeed follow the process $\SRow^*(t)$. Essentially, it will happen as long as the super-row is safe.

\begin{claim} \label{cl:srowSafe}
Suppose super-row $i$ is in state $\SRow^*(t)$ at some time $t\geq t_0$, and it is safe between time $t$ and time $t'> t$. Then at time $t'$ it is in state $\SRow^*(t')$.
\end{claim}

Now we can formulate our inductive hypothesis, i.e. specify what is state $\IndHyp_n$. We define it by specifying separately the state of each super-row: $\IndHyp_n$ is the state of our construction such that each super-row $i$ is in state
	$\SRow^*(t_0 + i_{n-i}\,\DeltaT)$.
Note that the sequence
	$\{\IndHyp_n:\,n\in\N\}$
has period \BigPeriod, as required.

Finally, we are ready to state and prove the theorem that our construction exhibits oscillations.

\begin{theorem}\label{thm:oscillations}
Suppose we initialize our construction at time $t_0$ in state $\IndHyp_0$. Then at time $t_0+\BigPeriod$ it is in the same state.
\end{theorem}

We formulate the proof as the next subsection.

\subsection{Full construction: proof of the main theorem}
\label{sec:mainThm}

It suffices to prove the inductive step. Let us assume that for some $n\in\N$, at time
	$t_1 := t_0+n\,\DeltaT$
our construction is in state $\IndHyp_n$, i.e. each super-row $i$ is in state
	$\SRow^*(t_0 + i_{n-i}\,\DeltaT)$.
We need to prove that at time
	$t_1+\DeltaT$
it is in state $\IndHyp_{n+1}$. Equivalently, we need to prove that super-row $n+1$ is in state $\SRow^*_0$, and any other super-row $i$ is in state
	$\SRow^*(t_0 + i_{n-i}\,\DeltaT + \DeltaT)$.
In short, we will get the former via replenishing processes (one for each row in super-row $n+1$), and the latter via Claim~\ref{cl:srowSafe}.

For simplicity let us assume that
	$\Nsup(\ConnProfile^*)\leq n<\Nsup$.
Then at time $t_1$ each super-row $i\leq n$ is in state
	$\SRow^*(t_1-i\,\DeltaT)$,
and super-row $n+1$ is in state
	$\SRow^*(t_0+ (\Nsup-1)\,\DeltaT)$.
Note that by the choice of parameter \Nsup\ at time $t_1$ all rows in super-row $n+1$ are max-stable.

Let us consider row $\mathcal{R}$ of super-row $n+1$. Let $Q$ and $D$ be the corresponding replenishing and decelerating servers, and let
	$t_2 := \tau_Q-\DeltaT^*$.
Note that $t_2>t_1$. Indeed, by~\refeq{eq:DeltaTprop}
\begin{eqnarray*}
\DeltaT^*  &\geq& (L_1+1)\,\TP = t_0\\
	t_2 &\geq& (n+1)\,\DeltaT + \TP - \DeltaT^* \\
	   &=& (n\,\DeltaT+t_0) + (\DeltaT^* - t_0) + \TP\\
	   &\geq& t_1+\TP.
\end{eqnarray*}	

\begin{claim}
At time $t_2$ our construction is $Q$-synchronized.
\end{claim}

\begin{myproof}
We need to check the  conditions in the definition~\refeq{eq:defnSynch}. Indeed, by the induction hypothesis at time $t_1$ row $\mathcal{R}$ is max-stable, each vertical session $V_{ik}\in V_Q$ is max-stable, and the corresponding super-row $i\leq n$ is in partial state $\SRow(t_1-i\,\DeltaT)$. Therefore by Claim~\ref{cl:srow}(c) at time $t_2$ it is in partial state $\SRow(t_2-i\,\DeltaT)$. Note that $t_2 = \tau_{ik}$ since the connectivity profile \ConnProfile\ is synchronized. It follows that at time $t_2$ super-row $i$ is in partial state $\SRow(\tau_{0k})$.

By Lemma~\ref{lm:safe}(a) during time interval $[t_1;\, t_2]$ session $V_{ik}$ stays safe. Since this happens for all sessions $V_{ik}\in V_Q$, during this time interval servers $Q$ and $D$ are empty, so row $\mathcal{R}$ stays max-stable. Moreover, by Lemma~\ref{lm:safe}(b) between times $t_1$ and $t_2$ super-row $i$ stays open for $V_{ik}$, so all servers on the flow path of $V_{ik}$ are open for $V_{ik}$, so it stays max-stable. We have checked all conditions in definition~\refeq{eq:defnSynch}.
\end{myproof}

By Lemma~\ref{lm:repl} it follows that at time $\tau_Q$ row $\mathcal{R}$ is in partial state
	$\Row(\TP)$,
each session $V_{ik}\in V_Q$ is happy, and moreover row $\mathcal{R}$ is open between times $t_2$ and $\tau_Q$. We use this to prove the following two claims.

\begin{claim}\label{cl:rowState}
At time $t_1+\DeltaT$ replenishing server $Q$ and decelerating server $D$ are empty, and row $\mathcal{R}$ is in partial state $\Row(t_0 - j\,\Deltat)$,
where $j$ is the row number of $\mathcal{R}$.
\end{claim}

\begin{myproof}
At time $\tau_Q$ all vertical sessions $V_{ik}\in V_Q$ are happy. It follows that they are safe and servers $Q$ and $D$ are empty. By Lemma~\ref{lm:safe}(a) these sessions are safe during time interval
	$[\tau_Q;\, t_1+\DeltaT]$.
Therefore at time $t_1+\DeltaT$ servers $Q$ and $D$ are empty.

Since at time $\tau_Q$ row $\mathcal{R}$ is in partial state $\Row(\TP)$, by Claim~\ref{cl:row}(c) at time $t_1+\DeltaT$ this row is in partial state $\Row(t)$, where
	$ t = \TP + t_1+\DeltaT - \tau_Q = t_0 - j\,\Deltat. $
\end{myproof}

\begin{claim}\label{cl:rowOpen}
Row $\mathcal{R}$ is open between times $t_1$ and $t+\DeltaT$.
\end{claim}

\begin{myproof}
We already proved that row $\mathcal{R}$ stays max-stable, hence open, between times $t_1$ and $t_2$. By Lemma~\ref{lm:repl} it is open between times $t_2$ and $\tau_Q$. Finally, at any time
	$t\in [\tau_Q,\, t_1+\DeltaT]$
this row is in partial state $\Row(t')$, $t'\leq t_0$,
and therefore it is open by~\refeq{eq:bound-k}.
\end{myproof}

Since Claim~\ref{cl:rowState} holds for all rows in super-row $n+1$, at time $t+\DeltaT$ this super-row  is in partial state $\SRow_{k_0}$. We need some more work to go from statements about partial states of super-rows to statements about their states.

\begin{claim} \label{cl:srowOpen}
Let $V$ be a vertical session $V$ in super-row $n+1$. Then between time $t_1$ and $t_1+\DeltaT$ (a) this super-row is open for $V$, and (b) session $V$ is safe.
\end{claim}

\begin{myproof}
Part (a) follows by applying Claim~\ref{cl:rowOpen} to every row in super-row $n+1$. Part (b) follows from part (a) since by induction hypothesis at time $t_1$ session $V$ is max-stable.
\end{myproof}

Say a replenishing server is \emph{essential} if it lies in super-row $n+1$. Say a vertical session $V$ is \emph{essential} if $V\in V_Q$ for some essential replenishing server $Q$. We show that non-essential vertical sessions and replenishing servers are safe as far as we are concerned.

\begin{claim}\label{cl:non-essentialVik}
For each non-essential vertical session $V_{ik}$, intervals $[t_1;\, t_1+\DeltaT]$ and
	$[\tau_{ik};\, \tau_{ik} + \DeltaT^*]$
are disjoint.
\end{claim}

\begin{proof} Suppose $V_{ik}\in V_Q$ for some replenishing server $Q$ in row $j$ of super-row $l\neq n+1$. If $l\leq n$ then
$$ \tau_{ik} + \DeltaT^* = l\,\DeltaT + j\,\Deltat
	<n\,\DeltaT + t_0 = t_1.
$$
If $l\leq n+2$ then
	$ \tau_{ik} > (n+2)\,\DeltaT - \DeltaT^* > t_1 + \DeltaT$.
\end{proof}

\begin{claim} \label{cl:non-essential}
Between time $t_1$ and $t_1+\DeltaT$, all non-essential vertical sessions and replenishing servers are safe.
\end{claim}

\begin{myproof} By induction hypothesis this condition holds at time $t_1$. Suppose it fails at time
	$t\in (t_1, t_1+\DeltaT]$,
and let us assume this is the first time it fails in this time interval.

If a non-essential replenishing server $Q$ becomes unsafe at time $t$, then at some time $t'\in (t_1, t]$ some vertical session $V\in V_Q$ must have become unsafe. Therefore without loss of generality a non-essential vertical session $V_{ik}$ becomes unsafe at time $t\in (t_1, t_1+\DeltaT]$. Let us argue to the contradiction.

By Claim~\ref{cl:srowOpen}(b) session $V_{ik}$ cannot belong to super-row $n+1$.  By induction hypothesis and~\refeq{eq:NcolsProp}, at time $t_1$ either super-row $i$ is empty or it is in a well-defined partial state $\SRow(t_1-\tau_{i0})$. In the first case by Lemma~\ref{lm:safe+empty} super-row $i$ becomes unsafe at some time
	$t'\in (t_1, t)$,
contradiction. So we are in the second case. If $t-\tau_{i0} > t_*$ then by Lemma~\ref{lm:safe}(c) super-row $i$ becomes unsafe at some time
	$t'\in (t_1, t)$,
contradiction. Else by Lemma~\ref{lm:safe}(a) we must have
	$t\in[\tau_{ik};\, \tau_{ik} + \DeltaT^*]$,
which contradicts Claim~\ref{cl:non-essentialVik}.
\end{myproof}

\begin{claim}
At time $t+\DeltaT$ super-row $n+1$ is in state $\SRow^*_0$.
\end{claim}

\begin{myproof}
We know that at time $t+\DeltaT$ this super-row  is in partial state $\SRow_{k_0}$, and (by Claim~\ref{cl:rowState}) all replenishing and decelerating servers in this super-row are empty.

It remains to prove that at this time each vertical session $V$ in this super-row is max-stable. This is the case because it is max-stable at time $t_1$ by the induction hypothesis, and during time interval $[t_1;\, t_1+\DeltaT]$ all servers on the flow path of $V$ are open for $V$. Namely,
 during this time interval the corresponding replenishing and decelerating servers are open for $V$ by Claim~\ref{cl:non-essential}, and super-row $n+1$ is open for $V$ by Claim~\ref{cl:srowOpen}(a).
\end{myproof}

For each super-row $i\neq n+1$, note that by Claim~\ref{cl:srowOpen}(b) and Claim~\ref{cl:non-essential} it is safe between time $t_1$ and $t_1+\DeltaT$. At time $t_1$ it is in state
	$\SRow^*(t_0+i_{n-1}\,\DeltaT)$
by the induction hypothesis, so by Claim~\ref{cl:srowSafe} at time $t_1+\DeltaT$ it is in state
	$\SRow^*(t_0+i_{n-1}\,\DeltaT + \DeltaT)$
as required.

This completes the proof of Theorem~\ref{thm:oscillations} and therefore the proof of Theorem~\ref{thm:main}.

\section{Same damping parameter for all sessions}
\label{sec:sameDelta}

In this section we fine-tune the main theorem (Theorem~\ref{thm:main}) so that all sessions have the same damping parameter. This result is stated in the Introduction as Theorem~\ref{thm:sameDelta}. Here we prove it in a somewhat more general form.

We consider flow decrease functions of the form
\begin{equation}\label{eq:splittableSessions}
	 f_\delta (\ve, t) = \ve - h(\ve, \delta)\, f(t),
\end{equation}
where for any fixed $\delta\in \VALUESofDelta$ we have
\begin{equation}\label{eq:splittableSessionsLimit}
	\lim_{\,\ve\rightarrow 0+}\, h(\ve,\delta)/\ve = \infty.
\end{equation}
To make such functions satisfy axioms~\refASS{ass:increase}{ass:continuity} from Section~\ref{sec:flowControl}, let us impose some natural constraints on smoothness and monotonicity:
\begin{itemize}

\item $f(t)$ is continuous on $[0, \infty]$ and increases from $0$ to $\infty$,

\item  $h(\ve, \delta)$ increases in $\delta$ from $0$ to $\infty$, for any fixed $\ve\geq 0$,

\item $h(0, \delta) = 0$ for all $\delta\in \VALUESofDelta$,

\item $h(\ve, \delta)$ is differentiable in $\ve$, for any fixed
	$\delta\in \VALUESofDelta$.

\item $h(\ve, \delta)/\ve$ is decreasing in $\ve$, for any fixed
	$\delta\in \VALUESofDelta$.
\end{itemize}
The last of these conditions is motivated by~\refeq{eq:splittableSessionsLimit}. It is included specifically to ensure that $f_\delta$ is increasing in $\ve$ whenever it is well-defined, i.e. whenever it is non-negative; one can see it easily by differentiating $h(\ve, \delta)/\ve$ and~\refeq{eq:splittableSessions} with respect to $\ve$.

If a flow decrease function function satisfies all these conditions, let us call it \emph{splittable}.  This definition is motivated by the fact that for such functions we will be able to split any session with damping parameter $\delta$ into several parallel sessions with any given damping parameter $\delta'<\delta$. An example of such functions is given in~\refeq{eq:splittableExample}.

Now we can state our result as follows:

\begin{theorem} \label{thm:AD}
Suppose in Theorem~\ref{thm:main} the flow decrease function is splittable. Then we can choose the same damping parameter for all sessions. Moreover, there exists $\delta^*\in\VALUESofDelta $ with the following property: for any
	$\delta\in\VALUESofDelta$ such that $\delta\leq \delta^*$
we can choose the damping parameter to be $\delta$.
\end{theorem}

\begin{proof}
Recall that in our construction we have three types of sessions: simple, horizontal and vertical, with three different pairs of flow control parameters $(\epsS,\deltaS)$, $(\epsH,\deltaH)$ and $(\epsV,\deltaV)$.
Let $\delta^*$ be the smallest of the three damping parameters. We will transform our construction so that all sessions have damping parameter $\delta^*$. (The same transformation also works if we replace $\delta^*$ by any given smaller value.)

Let us focus on one type of sessions. Let $(\ve,\delta)$ be its flow control parameters. If $\delta = \delta^*$ then we do not need to do anything. Else, we will replace each session of this type with several parallel sessions with damping parameter $\delta^*$ and maximal injection rates
	$(\ve_1, \ve_1,\, \ldots,\, \ve_k)$
that sum up to $\ve$, where the rates $\ve_i$ are chosen so that
\begin{equation}\label{eq:replaceSessions}
\text{
	$f_\delta (\ve,t) = \sum_{i=1}^k f_{\delta^*} (\ve_i, t)$\;
	for all times $t\geq 0$.
}\end{equation}
Assume such rates exist; we will prove it later. Then if these $k$ new sessions all become blocked when they are sending at their respective maximal injection rate, they decrease their sending rate exactly as the original session did. It is crucial that in our construction each session becomes blocked only when it is at the maximal injection rate.

For the modified construction, we choose \Nsup, the number of super-rows, exactly as before, except now we want it to be large enough so that all \emph{new} sessions have time to build up their injection rate to the maximum values. Then the modified construction works exactly as the original one.

It remains to prove that for any pair $(\ve,\delta)$ there exist rates
	$(\ve_1, \ve_1,\, \ldots,\, \ve_k)$
that sum up to $\ve$ and satisfy~\refeq{eq:replaceSessions}. Indeed, whenever the rates are non-negative  and sum up to $\ve$, we have
	$$  \sum_{i=1}^k f_{\delta^*} (\ve_i, t)
		= \ve - f(t) \sum_{i=1}^k h(\ve_i, \delta^*).
	$$
If $\ve_i = \ve/k$ for each $i$, then
	$$ \textstyle
		\sum_{i=1}^k h(\ve_i, \delta^*) = k\, h(\ve_1, \delta^*)
		= \ve\, h(\ve_1, \delta^*)/\ve_1,
	$$
which is greater than $h(\ve, \delta)$ if $k$ is large enough. Let us choose the smallest such $k$; note that $k\neq 1$ since $\delta^*<\delta$.

Now let us perturb the rates so that the above sum is exactly equal to $h(\ve, \delta)$. With this goal in mind, let us define a one-dimensional function
	$$ H(\ve_1) := h(\ve_1, \delta^*) +
		\sum_{i=2}^k h\left(
			\frac{\ve-\ve_1}{k-1},\, \delta^*
		\right)
	$$
Then $H(\ve_1)$ is continuous on $[0; \ve]$, and
	$H(\ve/k)> h(\ve, \delta)$.
Moreover,
	$$ H(0) = \sum_{i=1}^{k-1} h\left(
		\frac{\ve}{k-1},\, \delta^*
	\right),
	$$
which is less than or equal to $h(\ve, \delta)$ by the choice of $k$. Therefore there exists $\ve_1\in [0, \ve/k]$ such that
	$H(\ve_1) = h(\ve, \delta)$,
as required. For this value of $\ve_1$ we define the other rates accordingly:
	$\ve_i = (\ve-\ve_1)/(k-1)$
for each $i\geq 2$. This completes the proof of the theorem.
\end{proof}

\section{Further research: the all-FIFO setting}
\label{sec:allFIFO}

We conjecture that Theorem~\ref{thm:main} extends to the case when all
servers must be FIFO. We have a promising preliminary result in this
direction, namely an all-FIFO version of the basic gadget in which
server $Q_2$ is strictly FIFO and does not give session $H$ priority
over session $V$.

We consider the basic gadget from Section~\ref{sec:gadget}, with one modification that all four servers are now FIFO. Refer to Table~\ref{tab:paramsGadget} for the list of all relevant parameters. As before, we will have vertical session(s) going through server $Q_2$. Since now horizontal fluid does mix with vertical fluid, we have to consider the vertical session(s) explicitly inside the gadget.


We will allow $L$ vertical sessions coming through server $Q_2$ (in the full construction, parameter $L$ should define \Nvert\ and \Nrows). We denote these sessions by $V_i$, $1\leq i\leq L$. We do not want to characterize the arrival pattern of sessions $V_i$ at this point. Instead, we shall simply place a bound on the arrivals.  In particular, we let $f_i(t)$ represent the rate at which fluid arrives at server $Q_2$ on session $V_i$ at time $t$, and let
	$f(t) = \sum f_i(t)$.
We shall assume that
\begin{eqnarray}
	\mu_2 &<& \mu_3+ \epsS^2 				\label{eq:mu} \\
	f(t) &\le& \epsS^2 \;\; \text{for all times $t$} 	\label{eq:vertFlows}
\end{eqnarray}

In general, the vertical sessions are supposed to enter server $Q_2$ right after they leave from a similar server from some other gadget. For this reason we need to assume that the rates $f_i$ may depend on $\mu_2$. To this extent, we will assume the following:
\begin{eqnarray}\text{
	the intergral $\int_0^T f(t) \, dt$ is continuous in $\mu_2$
}
\label{eq:continuity}
\end{eqnarray}
for any fixed time $T$.

\begin{lemma}\label{lm:phaseFIFO}
Consider the basic gadget under all assumptions in Lemma~\ref{lm:phase}, but define all servers to be FIFO. Then:

\begin{itemize}
\item[(a)] There exist parameters  $(\epsS, \deltaS,\mu_3, K, K')$ and times $(T_1,T_3,P)$ such that given
	(\ref{eq:mu}-\ref{eq:vertFlows})
the basic gadget functions as shown in Table~\ref{tab:phase}.

\item[(b)] Part (a) holds for any given $K\geq 20$, in which case we can choose
	$\epsS \in (5/3K; 20/K)$
and set $\mu_3 = 2\epsS/5$.

\item[(c)] Moreover, for any $K\geq 100$ and the corresponding choice of parameters $(\epsS, \deltaS, K')$, and any rates $f_i(t)$ satisfying
	(\ref{eq:vertFlows}-\ref{eq:continuity})
there exists
	$\mu_2\in [\epsS/60; \epsS/5]$
satisfying~\refeq{eq:mu} such that the maximal height of server $Q_2$ is $P/\cel{5P}$.

\end{itemize}
\end{lemma}

Note that in part (a) one choice of parameters
	$(\epsS, \deltaS,\mu_3, K, K')$
works for \emph{any} parameter $\mu_2$ satisfying~\refeq{eq:mu}, and \emph{any} arrival rates $f_i(t)$ satisfying~\refeq{eq:vertFlows}. The idea is that first we choose the above five parameters using part (a), and then fine-tune $\mu_2$ using (c). Part (b) could be useful to make sure that $\epsS$ is as small as desired.

Part (c) is important because in the all-FIFO setting the time separation between two consecutive rows should be equal to the maximal height of server $Q_2$; making this height equal to $P/b$, $b\in N$ makes row $i+b$  synchronized with row $i$.

\begin{myproof}
The proof of parts (ab) follows that of Lemma~\ref{lm:phase} almost word-by-word, with two modifications:
\begin{itemize}
\item In the second paragraph of Section~\ref{sec:basicGadgetProof} we need to note that the session $H$ traffic mixes with the ``vertical'' traffic in session $Q_2$. However, by~\refeq{eq:vertFlows} session $H$ fluid leaves server $Q_2$ at rate at least $1-\mu_2-\epsS^2$.  Since this is larger than $1-\mu_3$, we start to get a buildup in server $Q_3$.

\item For part (a) we need to observe that times $T_1$, $T_3$ and $\TP$ depend only on the parameters
	$(K,\epsS,\deltaS)$
and are independent of $\mu_2$ and the flow rates $f_i$. To see this for time $T_1$, note that for any time $t\leq T_1$, height $h_1(t)$ is given by
\begin{equation}
	h_1(t) = 1+ \int_0^t (d_4(t) - 1) \; dt, \label{eq:h1}
\end{equation}
where $d_4(t)$ is uniquely determined by system
	(\ref{eq:systemA}-\ref{eq:systemD}),
and therefore by parameters $(K,\epsS,\deltaS)$.

\end{itemize}

We need more work to prove part (c). Recall that we assume $K\geq 100$; accordingly, by part (b) we can choose $\epsS\leq 1/60$. We'll state explicitely where we need this condition.

We vary $\mu_2$ in the interval $[\epsS/60; \epsS/5]$, keeping all other parameters fixed; note that any such $\mu_2$ satisfies~\refeq{eq:mu} as long as $\epsS\leq 1/5$, so all of the previous analysis applies.

The maximal height of server $Q_2$ becomes a function of $\mu_2$; we denote this height by
	$H_2(\mu_2)$.
Intuitively, this is a continuous function. Let us quickly verify that this is indeed so. Recall that $T_1$ is fixed. For any time $t\in [0; T_1]$,
\begin{equation} \label{eq:h2}
h_2(t)	= \frac{1}{1-\mu_2} \int_0^t (\mu_2 + f(t)) \; dt,
\end{equation}
where $ f(t) = \sum_{i=1}^L f_i(t)$. Since
	$ H_2(\mu_2) = h_2(T_1)$,
it is a continuous function of $\mu_2$ as long as we assume~\refeq{eq:continuity}.

We claim that $1/\epsS \leq T_1 \leq 3/\epsS$. Indeed, at any time $t\in [0; T_1]$, server $Q_1$ drains at rate $1-d_4(t)$, which by \refeq{eq:DFour} lies between
	$\frac{\epsS}{2}/(1+\frac{\epsS}{2}-\mu_3)$
and
	$\epsS/(1+\epsS-\mu_3)$.
Hence the time $T_1$ at which server $Q_1$ drains satisfies,
$$ \frac{1}{\epsS} \leq \frac{1+\epsS-\mu_3}{\epsS} \leq T_1 \leq
	\frac{1+\frac{\epsS}{2}-\mu_3}{\frac{\epsS}{2}} \leq \frac{3}{\epsS},
$$
claim proved.

By~\refeq{eq:h2}, at any time $t\in [0; T_1]$ height $h_2(t)$ increases at rate that lies between
	$\mu_2/(1-\mu_2)$
and $(\mu_2+\epsS^2)/(1-\mu_2)$. The latter number is at most $2\mu_2$ whenever $\epsS\leq 1/60$ and $\mu_2\in [\epsS/60; \epsS/5]$, which matches our initial assumptions; the proof is a simple but tedious calculation which we omit. Therefore
	$$\mu_2/\epsS \leq H_2(\mu_2) \leq 6\mu_2/\epsS,$$
so, in particular, $H_2(\epsS/60) \leq 1/10$, and $H_2(\epsS/5) \geq 1/5$.

Finally, let us choose an integer $b=\cel{5P}$. Since $P\geq 1$, we have
	$P/b\in [1/10; 1/5]$.
Since $H_2(\mu_2)$ is continuous, there exists a $\mu_2\in [\epsS/60; \epsS/5]$ such that
	$H_2(\mu_2) = P/b$.
\end{myproof}

Unfortunately, we are unable to construct an example where the
replenishing servers are strictly FIFO.  The main difficulty is that
for a FIFO replenishing queue, when the vertical sessions are creating
a buildup of fluid a small amount of session $H$ fluid will be served.
Since we would like the session $H$ fluid to start traversing the row
only after all the vertical session fluid has left the replenishing
queue, this small amount of session $H$ fluid is served too early.

By choosing parameters appropriately we can ensure that the amount of
session $H$ fluid that is served early is arbitrarily small. However,
as long as the amount is nonzero it will affect the operation of all
the subsequent gadgets in the row. We conjecture that the effect will
be minimal and will not affect the final result. We believe that an
interesting open problem is to prove this conjecture.

\section{Conclusions}
\label{sec:conclusions}

In this paper we have shown that natural flow control schemes can
create oscillations when interacting with networks of queues.
These oscillations occur because the queueing dynamics affect the rate
at which data passes through a server and so the arrival rate of a
session at a server can be different than the external session injection
rate.  In particular, our example is always feasible in the sense that
the total injection rate of all sessions passing through a server is
never bigger than the service rate of that server.  This provides a
contrast with previous work showing oscillations of TCP.

A number of open questions remain. The first set of questions concerns elaborating our construction. As already mentioned, we would like to extend our example to the case in which all servers are FIFO. Also, we would like to extend Theorem~\ref{thm:sameDelta} ('same damping parameter for all sessions') to a wider class of flow control functions, most notably to multiplicative decrease. Lastly, we would like to accomodate TCP Vegas-type schemes (see~\cite{BrakmoP95,LowPW02,ChoeL03}) where the flow rate is increased/decreased at a rate of $1/(\Gamma_i(t))^2$ depending on
whether or not there is congestion on the path of flow $i$.  Here
$\Gamma_i(t)$ is the end-to-end-delay on the path of flow $i$.

On the other hand, we wonder what would be a good way to \emph{break} our construction. In particular, it would be nice to prove a complimentary convergence result for our model under minor restrictions on the initial state, the way we fine-tune parameters, or the underlying network topology.

Another area for future study is reducing the complexity of our
example. We note that much of this complexity arises from the fact
that we wish to recreate our initial conditions {\em exactly} so that
we can create a infinite sequence of oscillations.  We are curious if
there is a much simpler example in which it is possible to create a
large finite number of oscillations.  In particular, we wonder if it
might be possible to create such examples on more natural topologies.

Another open question relates to the ``stablility'' of our
oscillating example. That is, if we slightly perturb the state of the system, will the oscillations persist?  We would also like to
investigate if our example can be applied to networks utilizing
different scheduling disciplines.  For example, we wonder whether TCP-like flow
control can exhibit oscillations when interacting with a network of
servers that schedule flows according to the \emph{Generalized processor sharing} scheme (see~\cite{DemersKS90, ParekhG93, ParekhG94}). Another feature of our example is that it utilizes large buffers.
Whenever a queue builds up the participating sessions recognize
congestion and then reduce their injection rates.  It would be
interesting to know if our example can be adapted to a small buffer
scenario in which fluid is dropped when congestion occurs.

Lastly, we would like to investigate to best way to prevent
oscillations via joint flow control and network scheduling.  A
promising approach in this direction is the \emph{Greedy primal-dual
algorithm} of Stolyar~\cite{Stolyar05-gpd}.

\bibliographystyle{IEEEtran}
\bibliography{IEEEabrv,my_abrv,infocom06_full}

\appendices

\section{Continuity issues}\label{app:continuity}

Let us return to Section~\ref{sec:basicGadgetProof} and consider the integral in~\refeq{eq:hplus} as a function of $\ve$ that we denote $F(\ve)$. We show that it is continuous in $\ve$. We will use the notation from Section~\ref{sec:gadget}.


\begin{claim}\label{cl:continuity}
$F(\ve)$ is continuous on $[\ve_1, \ve_2]$.
\end{claim}

We will need some basic facts from two-variable calculus.

\begin{fact}\label{fact:composition}
For sets $S,S'\subset \R^2$, consider functions
	$f: S \rightarrow \R$ and $g: S' \rightarrow \R$.
Let $h(x,y) = f(x, g(x,y))$ and assume that this function is well-defined on $S'$, i.e. that
$$\text{
 	$(x, g(x,y)) \in S$ for any $(x,y)\in S'$.
}$$
If $g$ is continuous at some point $(x,y)\in S'$, and $f$ is continuous at the corresponding point
	$(x, g(x,y))\in S$,
then the composition $h$ is continuous at $(x,y)$.

\end{fact}

\begin{fact}\label{fact:inverse}
For a closed rectangle $S\subset \R^2$, let $x(\ve, t)$ be a function $S\rightarrow \R$ which is continuous on $S$ and increasing in $t$. Then writing $x_\ve (t) := x(\ve,t) $, the function
	$f(\ve, x):= x_\ve^{-1} (x)$
is continuous on the set
	$\{ (\ve,x):\,  x = x(\ve, t) \text{ and } (\ve,t)\in S  \}. $
\end{fact}

\begin{fact}\label{fact:integral}
Consider a compact set $S\subset \R^2$ such that for any point $(x,y)\in S$, the segment
	$\{x\}\times [0,y]$ lies in $S$.
Suppose a function $f:S\rightarrow \R$ is continuous on $S$. Then the integral
	$g(x,y) := \int_0^y f(x,y)\, dy  $
is continuous on $S$.
\end{fact}

Equipped with these facts, let us prove Claim~\ref{cl:continuity}. Recall that $\eps(t)$, the injection rate into session $H$ at time $t$, is really a function of two variables, time $t$ and parameter $\eps$. For clarity, let us denote $f(\eps, t) := \ve(t)$. This function is defined on the closed rectangle
	$R = [\ve_1; \ve_2]\times [0; 2K]$;
it is continuous by~\refAS{ass:continuity}.

By~(\ref{eq:systemA}-\ref{eq:systemB}) and plugging in $\mu_2 = 2\ve/5$ we have
\begin{eqnarray*}
 x_\ve(t) &:=& t + h_4(t) = t(1-2\ve/5) + \int_0^t f(\eps,t)\, dt.
\end{eqnarray*}
This is a function of two variables $(\ve,t)$; by Fact~\ref{fact:integral} it is continuous on $R$.

Consider the set
\begin{eqnarray*}
R^* 	&:=& \{\, (\ve, x): x = x_\ve(t) \text{ and } (\ve,t)\in S  \,\} \\
	& =& \{\, (\ve, x): \ve\in[\ve_1; \ve_2]  \text{ and } x\in [ 0; x_\ve(2K)] \,\}.	
\end{eqnarray*}
Let us treat $ x_\ve(t)$ as a (parameterized) function of one variable, $t$, and consider its inverse
	$x_\ve^{-1}(x)$.
This inverse is well-defined for all $(\ve,x)\in R^*$. We can view it as a function of these two variables; by Fact~\ref{fact:inverse} this function is continuous on $R^*$. By Fact~\ref{fact:composition}
it follows that
	$ g(\ve,x) := f(\ve, x_\ve^{-1}(x))$
is continuous on $R^*$, too.

Let us recall that the rate $d_4(t)$ depends on time $t$ and parameter $\ve$; accordingly, let us write
	$d_4(t) = d_4(\ve,t)$
Then we can re-write~\refeq{eq:systemD} as
\begin{eqnarray*}
	d_4(\ve,x) &=& \frac{1-2\ve/5}{g(\ve,x)+1-2\ve/5}.
\end{eqnarray*}
By Fact~\ref{fact:composition} this is also a continuous function on $R^*$. By Fact~\ref{fact:integral} it follows that the function
\begin{eqnarray*}
	F(\ve,x) &:=& \int_0^t d_4(\ve, x)\, dx
\end{eqnarray*}
is continuous on $R^*$. Finally, recall that
$$ \text{
	$F(\ve) = F(\ve, T(\ve))$, where $T(\ve) = K/(1-2\ve/5)$.
}$$
By Fact~\ref{fact:composition} $F(\ve)$ is continuous on $[\ve_1, \ve_2]$, as required. This completes the proof of Claim~\ref{cl:continuity}.

\end{document}